\documentclass{aa}

\usepackage{graphicx}	
\usepackage{amsmath}	
\usepackage{amssymb}	
\usepackage{booktabs}
\usepackage{epsfig}
\usepackage{tablefootnote}
\usepackage{epstopdf}
\usepackage{xcolor}
\usepackage[colorlinks=true,citecolor=magenta,linkcolor=blue]{hyperref}
\usepackage{auto-pst-pdf}
\usepackage{longtable, ltxtable, booktabs, supertabular}
\usepackage{float}
\usepackage{multicol}
\usepackage{comment}
\usepackage{txfonts}

\DeclareUnicodeCharacter{2212}{-}
\begin{document} 

   \title{Wavelength-resolved Reverberation Mapping of quasar CTS C30.10: Dissecting MgII and FeII emission regions}
   \titlerunning{Wavelength-resolved RM}
   \authorrunning{Prince et al.}


   \author{Raj Prince \inst{1}\thanks{E-mail: raj@cft.edu.pl}, Michal Zaja\v{c}ek\inst{2,1}, Bo{\.z}ena Czerny\inst{1}, Piotr Trzcionkowski\inst{1}, Mateusz Bronikowski\inst{1,3}, Catalina Sobrino Figaredo\inst{4}, Swayamtrupta Panda\inst{1,5,6}, Mary Loli Martinez–Aldama\inst{1,7},  Krzysztof Hryniewicz\inst{8}, Vikram Kumar Jaiswal\inst{1}, Marzena {\'S}niegowska\inst{5,1}, Mohammad-Hassan Naddaf \inst{1,5},  Maciej Bilicki\inst{1}, Martin Haas\inst{4},  Marek Jacek Sarna\inst{5}, Vladimir Karas\inst{9},     Aleksandra Olejak\inst{5}, Robert Przy\l uski\inst{10}, 
   Mateusz Ra\l owski\inst{11}, Andrzej Udalski\inst{12}, Ramotholo R. Sefako\inst{13}, Anja Genade\inst{13, 14}, Hannah L. Worters\inst{13}          
      }

   \institute{Center for Theoretical Physics, Polish Academy of Sciences, Al. Lotnik{\'o}w 32/46, 02-668 Warsaw, Poland\\
              \email{raj@cft.edu.pl; zajacek@cft.edu.pl; bcz@cft.edu.pl}
              \and
  Department of Theoretical Physics and Astrophysics, Faculty of Science, Masaryk University, Kotl{\'a}\v{r}sk{\'a} 2, 611 37 Brno, Czech Republic  
  \and
  Centre for Astrophysics and Cosmology, University of Nova Gorica, Vipavska 11c, 5270 Ajdov\v{s}\v{c}ina, Slovenia
\and
Astronomisches Institut Ruhr-Universitat Bochum, Universitatsstrae 150, D-44801 Bochum, Germany
  \and
  Nicolaus Copernicus Astronomical Center, Polish Academy of Sciences, ul. Bartycka 18, 00-716 Warsaw,
Poland
\and
Laborat{\'o}rio Nacional de Astrof{\'i}sica, R. dos Estados Unidos, 154 - Nac{\~o}es, Itajub{\'a} - MG, 37504-364, Brazil
    \and
 Departamento de Astronomia, Universidad de Chile, Camino del Observatorio 1515, Santiago, Chile 
 \and
National Centre for Nuclear Research, ul. Pasteura 7, 02-093 Warsaw, Poland
\and
         Astronomical Institute, Academy of Sciences, Bo\v{c}n{\'i} II 1401, CZ-14131 Prague, Czech Republic    
         \and
     Space Research Center, Polish Academy of Sciences, Bartycka 18A, 00-716 Warszawa
     \and
     Astronomical Observatory of the Jagiellonian University, Orla 171, 30-244 Krakow, Poland
     \and
Warsaw University Observatory, Al. Ujazdowskie 4, 88-478 Warsaw, Poland
     \and
South African Astronomical Observatory, PO Box 9, Observatory, 7935 Cape Town, South Africa
\and
University of Cape Town, Rondebosch, Cape Town, 7700
             }

   \date{Received \ldots, 2022; accepted \ldots, 2022}

 
  \abstract
   {We present the results of the reverberation monitoring aimed at MgII broad line and FeII pseudocontinuum for the luminous quasar CTS C30.10 (z = 0.90052) with the Southern African Large Telescope covering the years 2012-2021.}
   {We aimed at disentangling the MgII and UV FeII variability and the first measurement of UV FeII time delay for a distant quasar.}
   {We used several methods for time-delay measurements, and determined both FeII and MgII time delays as well as performed a wavelength-resolved time delay study for a combination of MgII and FeII in the 2700 - 2900 \AA ~ restframe wavelength range.}
   {We obtain the time delay for MgII of $275.5^{+12.4}_{-19.5}$ days in the rest frame, while for FeII we have two possible solutions of $270.0^{+13.8}_{−25.3}$ days and $180.3^{+26.6}_{−30.0}$ in the rest frame. Combining this result with the old measurement of FeII UV time delay for NGC 5548 we discuss for the first time the radius-luminosity relation for UV FeII with the slope consistent with $0.5$ within uncertainties.}
   {Since FeII time delay has a shorter time-delay component but lines are narrower than MgII, we propose that the line delay measurement is biased towards the BLR part facing the observer, with the bulk of the Fe II emission may arise from the more distant BLR region, one that is shielded from the observer.}


   \maketitle
%

\section{Introduction}


It is widely accepted now that the central engine of an active galactic nucleus (hereafter AGN) consists of the central supermassive black hole (SMBH) and the accretion disk around it \citep[see][for reviews]{krolik_book1999,50_years_book2012,2021bhns.confE...1K}. The recent observations by the Event Horizon Telescope (EHT) collaboration of the nearest jetted AGN, M87, has provided an elegant proof that the AGN carries a Kerr SMBH of $(6.5 \pm 0.7) \times 10^9\,M_{\odot}$ \citep{EHT2019_shadow,EHT2019}. This is also the case of extremely low-luminous systems, such as Sgr~A*, the closest galactic nucleus, where the bound orbits of S stars and dusty objects provided the evidence for the compact mass of $\sim 4\times 10^6\,M_{\odot}$ \citep{2016ApJ...830...17B,2017ApJ...837...30G,2017ApJ...845...22P,2018A&A...615L..15G,2020A&A...636L...5G,2020ApJ...899...50P,2020A&A...634A..35P}.

However, despite an increasing amount of knowledge, we have a much less detailed understanding of the properties of the plasma/gas and dust located further away from the SMBH, at a fraction of a parsec and more. This material is responsible for the characteristic broad emission lines coming from the Broad Line Region (BLR) as well as the infrared emission originating in the dusty/molecular torus \citep[see][for a review]{netzer2015ara}. Broad emission lines from the BLR are the most characteristic features in the optical and UV spectra of bright AGN of type I \citep{1943ApJ....97...28S,1959ApJ...130...38W,1963Natur.197.1040S} that are viewed close to the symmetry axis of the system, including quasars. For type II AGN, broad lines are not visible in direct unpolarized light due to the obscuration by the thick dusty molecular torus. However, they can be revealed in polarized emission thanks to the scattering \citep[type II AGN NGC1068 was the first such a case that revealed broad Balmer lines, see ][]{1985ApJ...297..621A}, which led to the unification scheme of AGN where different viewing angles reveal different structures of the nuclear engine \citep{1993ARA&A..31..473A,1995PASP..107..803U}.  

Generally, the velocity width of the broad emission lines varies from source to source, and the past studies suggest that it could be between $\sim$10$^3$ km s$^{-1}$ to $\sim$10$^4$ km s$^{-1}$ (\citealt{Schmidt1963}, \citealt{Osterbrock1986}, \citealt{Boroson1992}, \citealt{Sulentic2000}, \citealt{Shen2011}). The large emission-line widths in the BLR are caused by the cloud motion, specifically the Doppler broadening, while the responsible radiative process for the broad-line emission is apparently caused by the photoionization by the X-ray/UV radiation of the inner accretion disk, as implied by the significant correlation and the associated time delay of the emission-line light curve with respect to the changes in the irradiating continuum. The so-called reverberation mapping (RM) studies have successfully been performed by now for more than hundreds of objects \citep[e.g.][]{lyuty1975,Kaspi2000,Peterson2004,Bentz2013,2018NatAs...2...63M,Grier_2017,Du2018,Oz_DES_2021}.  

The BLR is basically unresolved, apart from the most recent measurements in the near-infrared domain ($K$ band, $2.2\,{\rm \mu m}$) performed for three AGN (3C273,  IRAS 09149-6206, NGC 3783) with the infrared instrument GRAVITY at the Very Large Telescope Interferometer \citep{GRAVITY2018,2019Msngr.178...20A,GRAVITY2020,GRAVITY2021}, thanks to which near-infrared broad hydrogen lines were spatially resolved. These spatially resolved BLR detections confirmed that the BLR is best represented by a thick disc system that rotates around the central source under the influence of the central SMBH. Hence, the GRAVITY observations have justified the RM method for studying the dynamics of BLR and the SMBH. This technique has extensively been employed in active galactic nuclei (AGN) to measure the time lags between the two causally connected light curves. The measured time lags can be directly linked to the physical size of the system via the speed of light. There are three types of RM, namely BLR-RM, X-ray-RM, and the continuum RM mainly seen in AGN (\citealt{Cackett2021}). It was first proposed by \citet{Blandford1982} and \citet{Peterson1993} and later it has been widely used to estimate the size of the BLR, accretion-disk size and the structure as well as the SMBH mass in the AGN and quasars (\citealt{Kaspi2000}, \citealt{Peterson2004},  \citealt{2018NatAs...2...63M}). Recently, it has been discovered that the time delay estimated from the RM can also be used to estimate the luminosity distance of the AGN which can eventually be used to constrain the cosmological parameters \citep{watson2011,haas2011,Czerny2013,martinez2019,zajacek2021,2021arXiv211200052K,khadka2021}.

Studies of the BLR line widths combined with the RM have clearly shown several important properties of the BLR \citep[see e.g.][]{Wandel1999,2009NewAR..53..140G,Li_YR_2013,pancoast2014,Grier_2017}: (i) a considerable stratification of the line-emitting material and (ii) a prevailing Keplerian motion, confined to the accretion-disc plane, but with an additional inflow/outflow or turbulent component. GRAVITY/VLT observations nicely confirmed the conclusion about the overall flatness of the BLR configuration deduced previously from the spectral/variability studies.

Simple time delay measurements of a single emission line do not give much information about the BLR structure, apart from the mean (effective) radius of the emission. More information comes from studies of many emission lines in a given source, and/or from velocity-resolved measurements \citep{DoneKrolik1996,Wandel1999,Bentz2010,Denney2010,Grier2012,Rosa_2015,Lu_2016,Pei_2017,derosa2018,Du2018,Xiao_2018,Zhang2019,Hu_2020,Horne2021, U2021}. These studies have been done so far for relatively nearby sources, for selected objects, including extensively monitored source NGC 5548. This method can reveal the velocity structure of the medium, and therefore it is most suitable not only for most reliable measurement of the time delays of specific lines and establishing the inflow/outflow pattern superimposed on the circular motion but also for the determination of time delays of broader pseudo-continua, such as the optical FeII and particularly for UV FeII pseudo-continuum, strongly overlapping with MgII line. The optical FeII time delay has indeed been measured in a few lower-redshift sources \citep{Bian2010,Barth2013,HuChen2015,Hu_2020,Zhang2019}.

Monitoring of more distant objects is in general less frequent but basic time delay measurements in distant quasars were performed in MgII line \citep{2006ApJ...647..901M,Shen2016,Czerny2019,Zajacek2020,zajacek2021,Lira2018,Homayouni2020,Oz_DES_2021} and CIV line \citep{2005ApJ...632..799P,2006ApJ...641..638P,2006ApJ...647..901M,2015ApJ...806..128D,2018ApJ...865...56L,2019MNRAS.487.3650H,2019ApJ...887...38G,2019ApJ...883L..14S,2021ApJ...915..129K,2022MNRAS.509.4008P}. In our previous paper about the luminous quasar CTS C30.10, we reported the long-term measurement of MgII emission \citep{Czerny2019}. The reverberation mapping result using various methods revealed a time delay of 562$^{+116}_{-68}$ days (rest frame) between the $3000\AA$ continuum and the MgII line variations in this source. This result suggests that the radius-luminosity relation derived from the MgII matches with the previous results from the H$\beta$ line. In addition, using the sample of 68 MgII quasars, we demonstrated that the scatter along the radius-luminosity relation is mostly driven by the accretion rate intensity \citep{2020ApJ...903...86M}.

The aim of the current paper is the first determination of time delay of UV FeII with respect to the continuum. Our study is based now on eight years of spectroscopic data for a quasar CTS C30.10 coming from the dedicated monitoring with the Southern African Large Telescope (SALT). We have determined the MgII time delay in this source earlier, on the basis of shorter data ($\sim$6 yrs) \citep{Czerny2019}. Since UV FeII and MgII decomposition may be biased by the choice of a template, we perform a wavelength-resolved analysis for this source.  



The velocity-resolved spectroscopy is an important tool, and can be used to explore the relation between the emission line variations and their velocity information. Eventually, it can also be used to estimate the mass of the central SMBH. In the past, this method has been applied to more than 35 AGNs by various authors (\citealt{Bentz2010}; \citealt{Denney2010}; \citealt{Grier2012}; \citealt{Du2018}; \citealt{derosa2018}; \citealt{Xiao_2018}; \citealt{Zhang2019}; \citealt{Hu_2020}; \citealt{U2021}).

The paper is structured as follows. In Section~\ref{sec_observations_data} we describe spectroscopic and photometric data and their reduction. The determination of the mean and the RMS spectra, the construction of MgII, FeII, and wavelength-resolved light curves is covered in Section~\ref{sec_measurements}. Subsequently, in Section~\ref{sec_results}, we analyze the continuum and the emission-line variability, and we present the MgII and the FeII time delays as well as the wavelength-resolved reverberation mapping of the MgII$+$FeII complex. In Section~\ref{sec_discussion}, we discuss the implications for the BLR kinematics and we show the updated MgII radius-luminosity relation as well as the first construction of the UV FeII radius-luminosity relation. Finally, we summarize the main conclusions in Section~\ref{sec_conclusions}.

\section{Observations and Data reduction}
\label{sec_observations_data}
The source CTS C30.10 is a bright (V$=17.2$ mag, NED) quasar in the southern part of the sky identified under the Calan-Tololo Survey (\citealt{Maza1988, Maza1993}). It is located at the redshift of $z$ = 0.90052  (\citealt{Modzelewska2014}) with R.A.=  04$^{\rm h}$47$^{\rm m}$19.9$^{\rm s}$, decl. = -45$^{\rm d}$37$^{\rm m}$38.0$^{\rm s}$ (J2000.0). We have monitored this quasar since December 2012 and the last observation was done on 2021 March 25. The long-term photometric and spectroscopic data have been used in this study. In almost nine years of observations, the source has been visited 36 times by the SALT telescope, and therefore we have 36 observations. The observation number 6 was done on 2014 August 17 and identified as an outlier and eventually was dropped from the further study. Therefore, the further study presented in this work is based on 35 SALT visits.  

\subsection{Spectroscopy}
The spectroscopic observational setup is similar to what is described in \citet{Czerny2019}. Here, we have 36 observations, each observation consist of two observing blocks with almost 800 s exposure.
The details about the first 26 observations are provided in \citet{Czerny2019} (Table-1)\footnote{ \href{https://iopscience.iop.org/article/10.3847/1538-4357/ab2913/pdf}{Table-1}}. In this work, we present only the later observations in Table~\ref{tab:SALT_data}. The observation number 18 in \citet{Czerny2019} was removed because of being a strong outlier. In this study, we have properly calibrated the observation 18 and it is turnout to contain reliable data and hence used in this work. However, the observation number 6 is still outlier (due to very poor weather conditions) and has not been considered in this work. The reduction of raw SALT data is done by SALT telescope staffs, using the standard pipeline \citep{SALT2010SPIE}, and the description about the further procedure of data reduction is provided in \citet{Czerny2019} in details. To correct for the vignetting effects in the SALT spectra a proper calibration with the use of a standard star is done. The detail description of the procedure is can be found in the Section 2.1 of the \citet{Modzelewska2014}.

\subsection{Mg$II$ line fitting}
\label{sect:MgII_method}
The reduced and calibrated spectra has been model with all possible components including continuum power-law (disk emission), FeII pseudo-continuum, and two components of MgII emission line. The FeII and MgII emission lines are expected to produce in the BLR. The spectra covers the wide range of wavelengths from 2700$-$2900 {\AA} in the rest frame of the source. As suggested by \citet{Modzelewska2014}, the FeII pseudo-continuum was modeled by the empirical template d12-m20-20-5 provided by \citet{Bruhweiler2008} with cloud number density of 10$^{12}$ cm$^{-3}$, microturbulent velocity of 20 km s$^{-1}$, and the flux of hydrogen-ionizing photons above 13.6 eV is assume to be 10$^{+20.5}$ cm$^{-2}$s$^{-1}$. 
The FeII template was further convolved with a Gaussian profile of width 900 km s$^{-1}$ considering the broadening in the FeII lines. 

\begin{figure}
    \centering
    \includegraphics[scale=0.45]{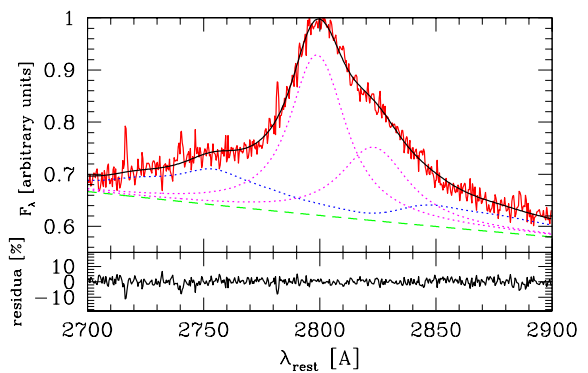}
    \caption{Observation 36 from SALT telescope (red) and the model (black). The green dashed line shows the underlying continnum from the accretion disk, blue dotted line is the FeII pseudo-continuum, and dotted magenta lines represent the two kinematic components of MgII line.}
    \label{fig:MgII_decomposition}
\end{figure}

Detailed modeling of MgII components are discussed in \citet{Modzelewska2014}. They have tried various ways to fit the MgII emission lines including single component, two separate emission components, and the single component with the absorption. Here, we have fitted the MgII emission lines with two different components assuming both components are described by Lorentzian shape profile. We also tried the Gaussian profile but the $\chi^2$ values were high compared to Lorentzian profile. An exemplary spectrum fitted with power-law, FeII, and MgII components is shown in Figure ~\ref{fig:MgII_decomposition}. 

\begin{table*}
  \caption{the results of the data fitting to SALT spectroscopy, starting from obs. 27. Earlier data are in \citet{Czerny2019}.}
  \label{tab:SALT_data}
  \centering                          
  \begin{tabular}{l r r r  r  r  r  r  r }        
    \hline\hline
    Obs. & JD  & EW(Mg II) & err+ & err- & EW(Fe II) & err+  & err- \\
    no.& - 2 540 000 &  [\AA] & [\AA] & [\AA] & [\AA] & [\AA] & [\AA] \\ 
    \hline
    27 & 8498.4492 & 23.37  &   0.37 &    0.36 &  6.91  &    0.63   &   0.63 \\
    28 & 8724.5739 & 26.81  &   0.34 &    0.34 &  11.18  &    1.07  &   1.06 \\
    29 & 8762.4741 & 25.72  &   0.49 &    0.44 &  10.87  &    0.83  &   0.87 \\ 
   30 & 8821.3060  & 27.93   &   0.54 &    0.52 &  12.47 &    0.97  &   0.99 \\ 
   31 & 8852.4699  & 26.82   &   0.56 &    0.58 &  11.19 &    1.06  &   1.08\\  
   32 & 9075.6137  & 28.26   &   0.42 &    0.39 &  12.31 &    0.702 &   0.69 \\  
   33 & 9116.5043  & 28.10   &   0.37 &    0.35 & 11.46  &    0.69  &   0.70\\
   34 & 9235.4390  & 28.60   &   0.66 &    0.63 & 12.78 &     1.09  &   1.10\\
   35 & 9291.2792  & 28.89   &   0.51 &    0.49 & 11.81 &     0.89  &   0.87\\
   36 & 9298.2681  & 30.28   &   0.62 &    0.56 & 11.73 &     1.12  &   1.11\\ 
   \hline \hline
  \end{tabular}
\end{table*}

\begin{table}[]
    \centering
     \caption{Estimated flux in MgII and FeII along with their errors. All the values and the errors are in units of 10$^{-14}$ erg s$^{-1}$cm$^{-2}$. As we pointed out earlier Obs no. 6 is an outlier in our data set. The FeII in this study is measured between 2700 to 2900 $\AA$ rest-frame.}
    \begin{tabular}{c|cc|cc}
    \hline \hline
    Obs no. & F(MgII) &err & F(FeII) &  err\\
\hline
 1 & 2.997 & 0.033 & 0.928 & 0.010 \\
 2 & 2.964 & 0.032 & 0.904 & 0.010 \\
 3 & 2.909 & 0.032 & 1.079 & 0.012 \\
 4 & 2.897 & 0.032 & 1.231 & 0.013 \\
 5 & 2.864 & 0.031 & 1.097 & 0.012 \\
 6 & 3.670 & 0.040 & 1.565 & 0.017 \\
 7 & 2.890 & 0.032 & 1.036 & 0.011 \\
 8 & 2.840 & 0.031 & 0.951 & 0.010 \\
 9 & 2.352 & 0.026 & 1.105 & 0.012 \\
 10 & 2.830 & 0.031 & 1.052 & 0.011 \\
 11 & 2.868 & 0.031 & 0.958 & 0.010 \\
 12 & 2.667 & 0.029 & 0.987 & 0.011 \\
 13 & 2.804 & 0.030 & 1.056 & 0.011 \\
 14 & 2.592 & 0.028 & 0.862 & 0.010 \\
 15 & 2.761 & 0.030 & 1.065 & 0.012 \\
 16 & 2.608 & 0.028 & 1.056 & 0.012 \\
 17 & 2.345 & 0.026 & 0.447 & 0.005 \\
 18 & 2.421 & 0.027 & 0.884 & 0.010 \\
 19 & 2.714 & 0.030 & 0.928 & 0.010 \\
 20 & 2.490 & 0.027 & 0.712 & 0.008 \\
 21 & 2.461 & 0.027 & 0.802 & 0.009 \\
 22 & 2.664 & 0.029 & 1.085 & 0.012 \\
 23 & 2.719 & 0.030 & 1.025 & 0.011 \\
 24 & 2.648 & 0.029 & 1.074 & 0.012 \\
 25 & 2.602 & 0.028 & 0.959 & 0.010 \\
 26 & 2.410 & 0.026 & 0.820 & 0.009 \\
 27 & 2.555 & 0.028 & 0.840 & 0.009 \\
 28 & 2.942 & 0.032 & 1.508 & 0.016 \\
 29 & 2.777 & 0.030 & 1.294 & 0.014 \\
 30 & 2.868 & 0.031 & 1.427 & 0.016 \\
 31 & 2.681 & 0.029 & 1.246 & 0.014 \\
 32 & 2.770 & 0.030 & 1.348 & 0.015 \\
 33 & 2.919 & 0.032 & 1.342 & 0.015\\
 34 & 2.819 & 0.031 & 1.375 & 0.015 \\
 35 & 2.798 & 0.030 & 1.249 & 0.014 \\
 36 & 2.905 & 0.032 & 1.208 & 0.013 \\
 \hline \hline
    \end{tabular}
    \label{tab:my_label}
\end{table}

The total equivalent width (EW) of the lines and their errorbars for the first 26 observations are presented in \citet{Czerny2019} and rest of the observations are shown in the Table~\ref{tab:SALT_data} of this work. Detailed descriptions of the parameters can be found in \citet{Czerny2019}. 

\subsection{Photometry}
\begin{figure}
    \centering
    \includegraphics[scale=0.35]{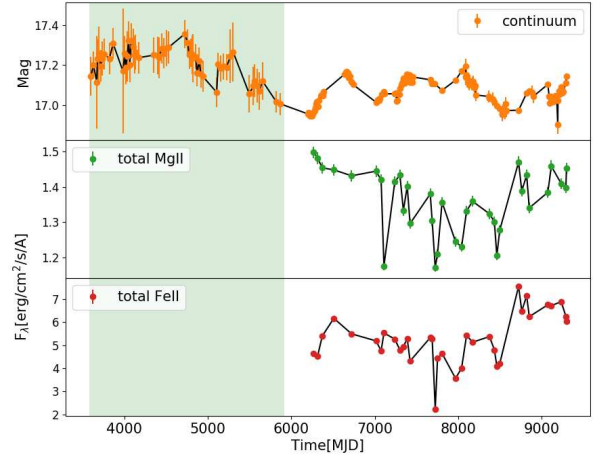}
    \caption{Long-term photometric light curve (upper panel), including CATALINA measurements (light green shaded region) from \citet{Czerny2019}, and total MgII (middle panel) and FeII (lower panel) lightcurves. Photometric observations are in magnitudes (panel 1) however, MgII and FeII fluxes are in units of 10$^{-16}$ and 10$^{-17}$ erg s$^{-1}$ cm$^{-2}$. In this study only non-shaded region is used. }
    \label{fig:total_Mg_Fe}
\end{figure}

The photometric observations of the source have been done with various telescopes across the globe. Our aim was to have the photometric observations close in time to spectroscopy by SALT, and hence in this regard we alerted many telescopes. Early part of the photometry was already described in \citet{Czerny2019}, where the photometric points from 4 telescopes were used (CATALINA survey for the very early observations prior to our monitoring, and later OGLE, SALTICAM SALT, and BMT (Observatorio Cerro Armazones, OCA). Here we include new data from SALTICAM SALT and BMT, as well as the data from four additional telescopes: Las Cumbres - Siding Spring Observatory (SSO) in Australia, SAAO (Lesedi Telescope), Las Cumbres - Cerro Tololo Inter-American Observatory (CTIO) and Las Cumbres - SAAO. There were small systematic shifts between the data from different telescopes so we applied a grey shift correction. Since some of the newly included photometry partially covers the time span covered by photometry given in \citet{Czerny2019}, we include the full photometry (apart from the CATALINA data) in Table~\ref{tab:photometry}. The resulting photometric curve is relatively smooth, but with clear variability pattern (see Figure~\ref{fig:total_Mg_Fe}). 

\section{Measurements}
\label{sec_measurements}
\subsection{Mean and RMS spectra}
To characterize the spectral behavior and the amplitude variation at different wavelengths, we also plot the mean and the root mean square (rms) spectra of the source. The mean and the rms spectra are defined as,

\begin{equation}
 \bar{F}_{\lambda} = \frac{1}{N} \sum_{i = 1}^{N} \textit{F}_{\lambda}^{i}
\end{equation}
and 
\begin{equation}
    S_{\lambda} = \left[ \sum_{i = 1}^{N} (\textit{F}_{\lambda}^{i} -  \bar{F}_{\lambda})^2   \right]^{1/2}
\end{equation}
where, 
$\textit{F}_{\lambda}^{i}$ is the $i$-th spectrum and $N$ is the number of spectra. 

\subsection{MgII and FeII lightcurves}

The lightcurves for the MgII and FeII were created as in our previous papers \citep{Czerny2019,Zajacek2020,zajacek2021}. The decomposition of each spectrum was done as described in Section~\ref{sect:MgII_method}. Each spectrum was then normalized using the photometric data since SALT spectroscopy does not allow for reliable spectrophotometric measurements directly. 
The final computation of the MgII and FeII flux was done by subtracting the power law component and the FeII or MgII component, correspondingly. The resulting lightcurves are shown in Figure~\ref{fig:total_Mg_Fe}, middle and bottom panel.

\subsection{Wavelength-resolved lightcurves}
We have followed the \citet{Hu_2020} method to divide the lightcurves according to the flux distribution in the flux rms spectrum. However, the \citet{Hu_2020} method was performed for H${_\beta}$ and in this paper we present for MgII and FeII emission.
After fitting the spectroscopic data with the full model, consisting of the power law continuum, FeII and MgII, we subtracted the power law component from the data in each data set. We then constructed the rms spectrum of the spectrum, and we divided the spectrum into seven bins of equal fluxes. We used less bins than in \citet{Hu_2020} since our data are of lower quality. 
We used the wavelength instead of velocity since we kept a combined contribution of MgII and FeII in the remaining spectrum due to considerable overlap of the two components. The separation of MgII and FeII is not unique, as we discussed in \cite{Zajacek2020} since it depends on the adopted template, and we actually aim at gaining additional insight into their separation directly from variability.

Thus defined wavelength bins were later used to create seven lightcurves from each original spectrum, again after subtraction of the best fitted power law and integrating each spectrum in the appropriate limits. 

\section{Results}
\label{sec_results}

Our lightcurve of CTS C30.10 is relatively long (over 8 years in the observed frame) in comparison with the time delay of 3 years in the observed frame, claimed in the previous paper \citep{Czerny2019} which allows for much better analysis. However, the overall source variability is still not much higher then before, which supports the view that frequency break in the power spectra of quasars is typically at 1-2 years \citep[e.g.][]{kozlowski2016,Stone2022}, and the variability amplitude rises much more slowly with extension of observing time. This saturation of the amplitude was already well seen in CTS C30.10 in the previous data (see Fig. 15 of \citealt{Czerny2019}).

\subsection{Variability}
The strength of the variability can be quantified by the excess variance ($\sigma_{XS}$), and the fractional rms variability amplitude, F$_{var}$ (\citealt{Edelson_2002}). The $\sigma_{XS}$ is the measurement of intrinsic variability in quasar, and estimated by correcting the total observed light curve with measurement errors. The F$_{var}$ is the square root of the $\sigma_{XS}$ normalized by the mean flux value.

The fractional variability is used to characterized the long-term variability in various bands. Its functional form and error on F$_{var}$ is taken from \citet{Vaughan_2003}.

\begin{equation}
F_{\rm var} = \sqrt{\frac{\sum_{i=1}^{N}((f_i -F)^2 - err_i^2}{F^2 (N-1)}},
\end{equation}
where F denotes the mean flux value, $f_i$ is the individual measurement, and err$_i^2$ is the error in the observed flux. The expression for the error on F$_{var}$ is provided in \citet{Prince_2019}.
We also estimated the point-to-point variability which tells about the variability at the shortest time scales. Considering the light curve is denoted by the f$_{i}$, where i= 1,2,3...N, the point-to-point variability is defined as ,
\begin{equation}
        F_{pp} = \sqrt{\frac{\sum_{i=1}^{N-1}(f_{i+1} - f_i)^2 - err_i^2 - err_{i+1}^2}{F^2 (N-1)}}. 
\end{equation}

The results are given in Table~\ref{tab:Fvar}. 
As expected the continuum shows less variability than the individual curves. The curve-1,2,3, and 7 have variability of more than 10\% in linear scale and curve-4,5, and 6 which is dominated by the MgII and FeII contribution show the variability below 10\%. The lower fractional variability is also noticed in total MgII and FeII emission consistence with the curve-4,5, and 6.
Further, to check the short scale variability, we estimated F$_{pp}$, for all the curves. For noisy individual curves F$_{pp}$ is of the same order as F$_{var}$. This confirms that the measurements are not dominated by the measurement errors since for the white noise F$_{pp}$ = 1.4 F$_{var}$, but variations are quite strong in the shortest timescales. This fast variability is not seen in the continuum, since for the continuum F$_{pp}$ is equal zero. Therefore, these fast variations are not a response to the continuum. Either emission lines show short timescale intrinsic variations, or we underestimate the measurement errors. Both effects are the potential sources of the problem, leading to a scatter in line-continuum relation. The poor correlation between the illuminating hard energy photons and the response of the line is frequently seen in many sources \citep[e.g.][]{Gaskell2021}, although best documented in the monitoring of NGC 5548 \citep{Goad2016, Gaskell2021}.

\begin{table*}[]
    \centering
    \caption{Variability amplitude for all the curves and the continuum used in this study.}
    \begin{tabular}{c|c|c|c|c|c}
    \noalign{\smallskip}
    \hline  \noalign{\smallskip}
  & &  \multicolumn{2}{c}{F$_{\rm var}$ } & \multicolumn{2}{c}{F$_{\rm pp}$ 
  } \\
  \noalign{\smallskip}
    \hline  \noalign{\smallskip}
 Rest frame Wavelength (\AA) &   Light curves & in linear scale($\%$)& in magnitude & in magnitude& in linear scale ($\%$) \\
    \noalign{\smallskip}
    \hline  \noalign{\smallskip}
2700.00 $-$ 2725.95 & Curve-1 &  17.00 $\pm$ 0.19  & 0.1922& 0.1961 &16.27$\pm$0.19 \\
2725.95 $-$ 2750.52 &Curve-2 &  14.88 $\pm$ 0.19   & 0.1704& 0.1818 & 15.32$\pm$0.19\\
2750.52 $-$ 2774.74 &Curve-3 &  13.88  $\pm$ 0.19  & 0.1573& 0.1564 & 13.11$\pm$0.18\\
2774.74 $-$ 2800.69 &Curve-4 &  7.64 $\pm$ 0.19    & 0.0854& 0.0903 & 7.95$\pm$0.18\\
2800.69 $-$ 2826.30 &Curve-5 &  7.68 $\pm$ 0.19    & 0.0846& 0.0840 & 7.49$\pm$0.18\\
2826.30 $-$ 2860.55 &Curve-6 &  8.97 $\pm$ 0.19    & 0.0985& 0.1069 & 9.61$\pm$0.18\\
2860.55 $-$ 2899.65 &Curve-7 &  14.10 $\pm$ 0.19   & 0.1583& 0.1655 & 14.29$\pm$0.19 \\
{\bf V-band} &   Continuum& 5.65 $\pm$ 0.16 & 0.0607 & 0.0 &0.0 \\
  MgII  & total & 6.63 $\pm$0.19 & 0.0727 & 0.0764 & 6.79$\pm$0.18  \\
  MgII & Comp-1 & 8.78$\pm$0.18  & 0.0959 & 0.1153 & 10.56$\pm$0.18  \\
  MgII & Comp-2 & 11.28$\pm$0.19 & 0.1227 & 0.1708 & 15.66$\pm$0.19 \\
  FeII & total & 20.61$\pm$0.19 & 0.2445 & 0.2614 & 19.91$\pm0.19$ \\
   \noalign{\smallskip}   \hline  \noalign{\smallskip}
    \end{tabular}
    \label{tab:Fvar}
\end{table*}

\subsection{Time delay measurement in total MgII and FeII }

\begin{table*}[]
    \centering
     \caption{Overview of the time-delay determinations for the total MgII and FeII total line emissions. The time delays are expressed in days with respect to the observer's frame, unless otherwise stated. The errors estimated for the mean are $1\sigma$ standard deviations, while for the time-delay peak the uncertainties consider the 30\% of the peak distribution. The median uncertainties express $16\%$ and $84\%$ percentiles of the distribution.}
    \begin{tabular}{c|c|c}
    \hline
    \hline
    Method  & MgII total [days] & FeII total [days] \\
    \hline 
    \smallskip
    ICCF (centroid)     & $381.06^{+79.96}_{-113.15}$  & $319.99^{+62.35}_{-50.40}$ \\ 
    ICCF (peak) & $383.50^{+73.32}_{-105.32}$  & $341.00^{+49.00}_{-51.77}$ \\
    \smallskip
    ICCF (max $r$) & $382.0$\,, $r=0.55$    & $340.0$\,, $r=0.65$   \\
    \hline 
    Javelin - peak (1 run) & $531.0^{+3.0}_{-6.6}$   & $504.0^{+18.3}_{-0.5}$\\
    Javelin(bootstrap, peak) &$529.0^{+24.4}_{-36.7}$ & $502.0^{+25.8}_{-51.8}$ \\
     Javelin(bootstrap, mean \& median) & $504.9^{+30.4}_{-28.1}$, $528.7^{+3.2}_{-88.2}$ & $494.3^{+25.5}_{-51.9}$, $503.3^{+30.8}_{-44.9}$ \\
    \hline
    $\chi^2$ (1 run)   & $535.6^{+12.0}_{-26.0}$   & $324.5^{+1.0}_{-4.0}$ \\
    $\chi^2$ (bootstrap peak)   & $539.0^{+25.0}_{-46.4}$   & $333.0^{+25.1}_{-25.6}$ \\
     $\chi^2$ (bootstrap mean \& median)   & $526.5^{+307.1}_{-86.7}$, $456.3^{+144.3}_{-118.1}$   & $360.9^{+201.3}_{-38.9}$, $343.2^{+43.0}_{-29.0}$ \\
    \hline
    von Neumann (1 run)   &    $512.0-514.0$       &    $512.0-514.0$           \\
    von Neumann (bootstrap peak) & $511.3^{+12.4}_{-35.6}$   & $511.3^{+21.6}_{-52.2}$ \\
    von Neumann (bootstrap mean \& median) &   $328.8^{+161.1}_{-543.9}$, $446.0^{+77.0}_{-335.0}$                 & $257.4^{+221.6}_{-607.4}$, $350.0^{+162.0}_{-473.0}$  \\
    \hline
    Bartels (1 run)    &    $512.0-514.0$                     &    $512.0-514.0$            \\
    Bartels (bootstrap peak)  & $511.3^{+21.8}_{-41.4}$    & $511.8^{+33.6}_{-52.9}$  \\
    Bartels (bootstrap mean \& median) & $351.6^{+187.2}_{-626.1}$, $487.5^{+35.5}_{-149.5}$    & $291.3^{+238.1}_{-602.8}$, $496.0^{+27.0}_{-614.5}$  \\
    \hline
    DCF (1 run)     & $527.5$   & $527.5$   \\
    DCF (bootstrap peak) & $527.0^{+31.7}_{-25.2}$ &  $527.0^{+23.6}_{-34.7}$  \\
    DCF (bootstrap mean \& median) & $431.8^{+40.3}_{-35.5}$, $397.5^{+130.0}_{-65.0}$ &  $461.9^{+26.5}_{-26.6}$, $527.5^{+0.0}_{-165.0}$  \\
    \hline
    zDCF (Maximum Likelihood peak \& full range)    & $291.3^{+190.2}_{-149.2}$\,,$(142.1, 481.5)$   & $ 353.6^{+77.6}_{-93.2}$\,,$(260.4, 431.2)$   \\
    \hline
    Mean time-delay peak - observer's frame  &  $523.5^{+23.6}_{-37.1}$   &  $513.0^{+26.2}_{-48.1}$ or $342.5^{+50.6}_{-57.0}$         \\
    Mean time-delay peak - rest frame &   $275.5^{+12.4}_{-19.5}$   &   $270.0^{+13.8}_{-25.3}$ or $180.3^{+26.6}_{-30.0}$                       \\
    \hline
    \end{tabular}
    \label{tab_totalMgII_FeII}
\end{table*}

\begin{figure}
    \centering
    \includegraphics[width=\columnwidth]{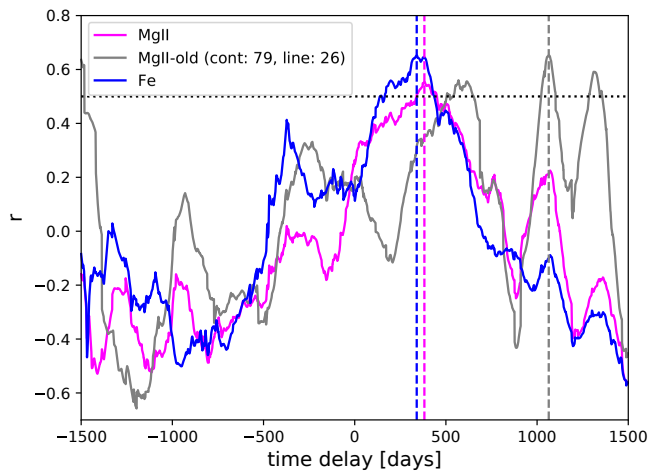}
    \caption{The interpolated cross-correlation function as a function of the time delay in the observer's frame for the total MgII (magenta line) and the total FeII emission (blue line). The dashed vertical lines mark the corresponding time-delay peak values. We see that for the FeII pseudocontinuum, the correlation coefficient at the peak value is larger than the correlation coefficient at the time-delay peak of the MgII line. The dotted horizontal line marks $r=0.5$. We also show the previous ICCF \citep{Czerny2019} when the peak time delay was at $\sim 1064$ days, see the gray solid line. Cont:79 and line:26 represents the number of observations used in \citep{Czerny2019}.}
    \label{fig_iccf_mg_fe}
\end{figure}

\begin{figure}
    \centering
    \includegraphics[width=\columnwidth]{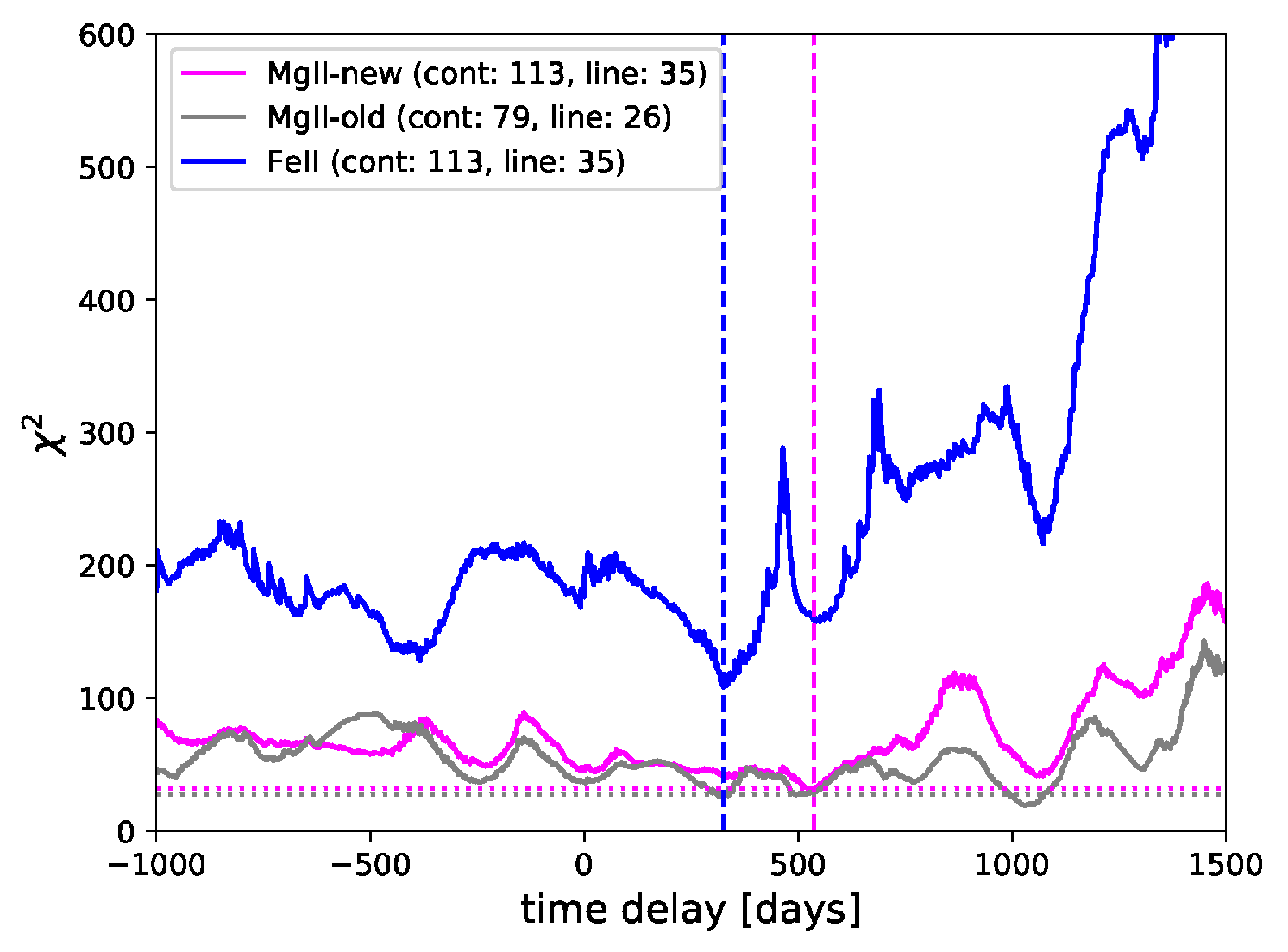}
    \caption{The $\chi^2$ value as a function of the time delay in the observer's frame for the MgII line emission (magenta line) as well as for the FeII line emission (blue line). The global $\chi^2$ minima for each line are depicted by the vertical dashed lines. The $\chi^2$ time-delay dependency for the previous, shorter continuum-MgII light curves \citep{Czerny2019} is shown as a gray line with the global minimum at the time delay twice as large as for the current data. Cont:79, 113 and line:26, 35 represent the number of observations.}
    \label{fig_chi2_total}
\end{figure}

\begin{figure}
    \centering
    \includegraphics[width=\columnwidth]{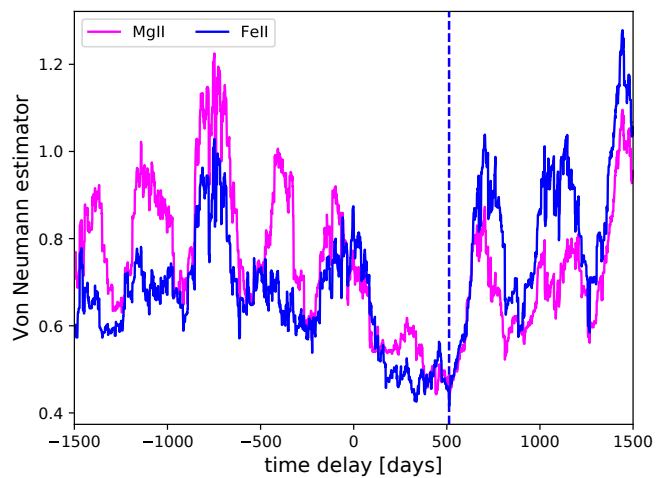}
    \caption{Von Neumann estimator as a function of the time delay in the observer's frame for both the total  MgII (magenta line) and FeII (blue line) emissions. The dashed vertical line marks the common global minimum of the von Neumann estimator at $\sim 512.0$ days.}
    \label{fig_vonneumann_whole}
\end{figure}

\begin{figure}
    \centering
    \includegraphics[width=\columnwidth]{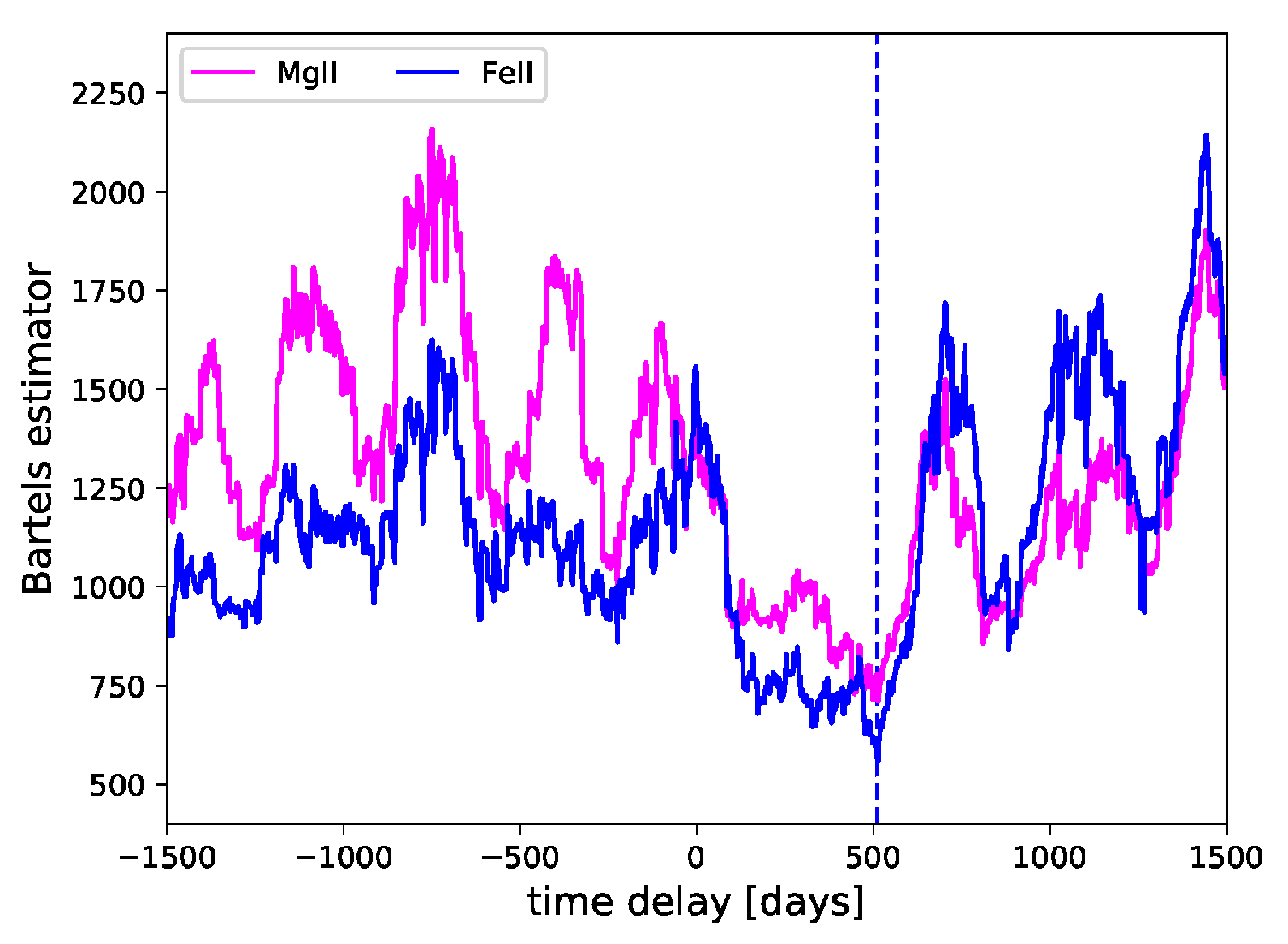}
    \caption{Bartels estimator as a function of the time delay in the observer's frame for both the total  MgII (magenta line) and FeII (blue line) emissions. The dashed vertical line marks the common global minimum of the Bartels estimator at $\sim 512.0$ days.}
    \label{fig_bartels_whole}
\end{figure}

\begin{figure}
    \centering
    \includegraphics[width=\columnwidth]{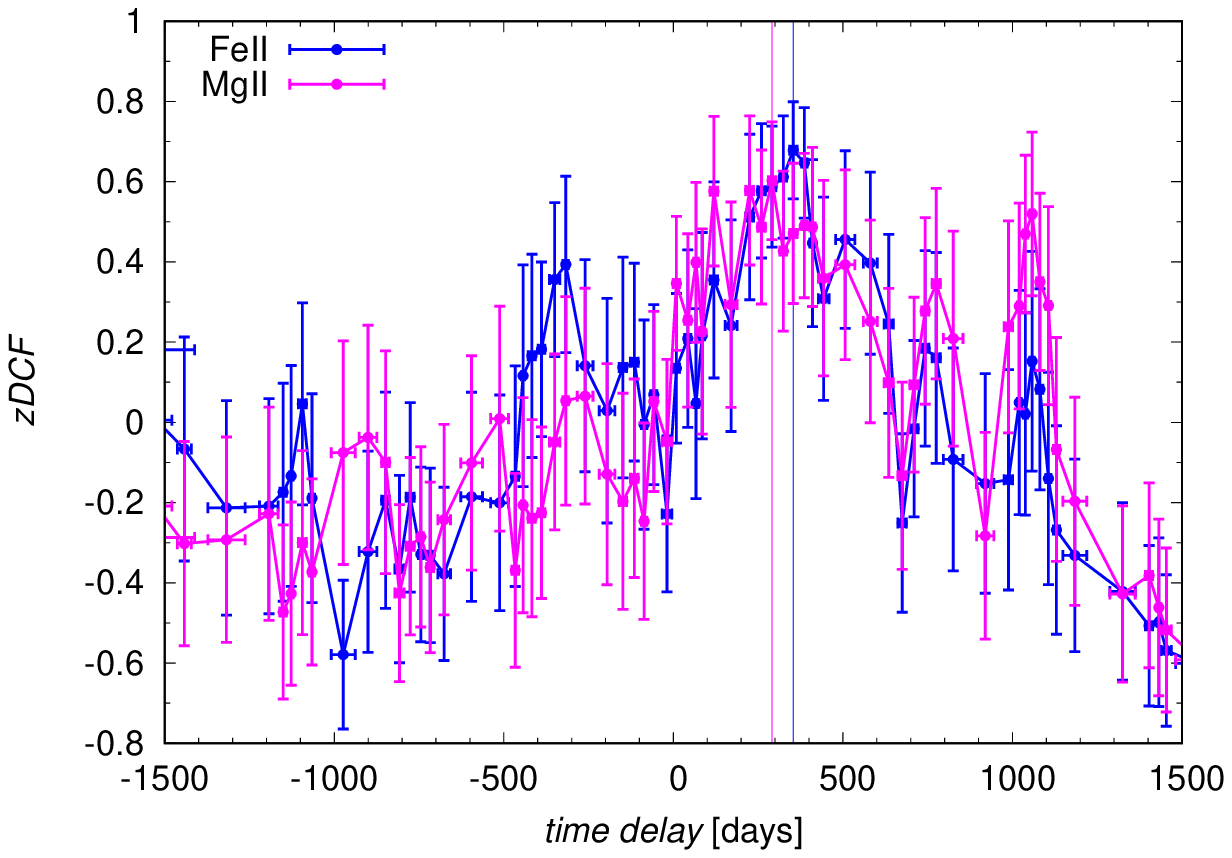}
    \caption{The zDCF correlation coefficient as a function of the time delay in the observer's frame for the total FeII (blue line) and the total MgII emission (magenta line). The vertical lines denote the peak values for each corresponding line.}
    \label{fig_zdcf}
\end{figure}

We apply several standard methods that are described in more detail in Appendix~\ref{subsec_timelag_methods}.
The time-delay evaluation using the standard interpolated cross-correlation function (ICCF) indicates a moderately longer peak time delay for the MgII broad line, $383.50^{+73.32}_{-105.32}$ days in the observer's frame, in comparison with the FeII pseudo-continuum, $341.00^{+49.00}_{-51.77}$ days (see Table~\ref{tab_totalMgII_FeII}). Interestingly, the correlation coefficient at the peak is higher for the FeII pseudo-continuum, $r=0.65$ versus $r=0.55$ for the MgII line, which is also visible in Figure~\ref{fig_iccf_mg_fe}, where we plot the ICCF as a function of the time delay for both lines. We stress that due to a large dataset (continuum points and line-emission measurements), the overall time-delay peak became smaller. Previously, in \citet{Czerny2019}, we reported the MgII time delay of $\sim 1050$ days in the observer's frame. We compare the previous ICCF and the current one in Figure.~\ref{fig_iccf_mg_fe}. We see that the shift of the best fit time delay is due to a change in the relative importance of the peaks in multipeak solution. We still see a trace of the 1050 days delay, but the shorter time delay has now higher significance.

The JAVELIN code, which models the continuum variability as a damped random walk, reveals a significant peak around $\sim 500$ days in the observer's frame for both the MgII line and the FeII pseudocontinuum. The peak time delay for the MgII emission is longer by $\sim 30$ days with respect to the FeII emission time delay, i.e. $530.0^{+25.4}_{-40.0}$ versus $500.0^{+25.2}_{-50.5}$ days, respectively, see Table~\ref{tab_totalMgII_FeII}, however, this is not a significant difference given the uncertainties. These peaks and their uncertainties were inferred from 100 bootstrap realizations based on the actual continuum and MgII and FeII line-emission light curves.   

The $\chi^2$ method shows a significant difference between the MgII and FeII line-emission time delays, 535.6 versus 324.5 days, respectively, see Table~\ref{tab_totalMgII_FeII}. In Figure.~\ref{fig_chi2_total}, we compare the $\chi^2$ dependency on the time delay for FeII (blue line) and MgII (magenta line) lines, which clearly depicts the shift for the MgII $\chi^2$ minimum towards a larger time delay. In addition, we compare the $\chi^2$ dependency between the older MgII and continuum data \citep{Czerny2019} and the current light curves. For a significantly larger number of continuum and line-emission data points, the MgII time delay gets smaller by approximately a factor of two. Based on the current datasets, we perform $10\,000$ bootstrap realizations, from which the MgII time delay is $539.0^{+25.0}_{-46.4}$ days that is by $\sim 200$ days longer than the FeII peak time delay of $333.0^{+25.1}_{-25.6}$ days. 

The estimators of data regularity/randomness (von Neumann and Bartels) indicate the minimum estimator value for the time delay of $\sim 511-512$ days in the observer's frame for the MgII line, see Table~\ref{tab_totalMgII_FeII}, where we inferred the peak and the mean time delays and the corresponding peak uncertainty based on 1000 bootstrap realizations for each estimator. Essentially the same best time-delay is also found for the FeII line. However, the mean time-delay value for the FeII line is smaller in comparison with the mean value for the MgII line, i.e. $273.0$ days vs. $313.8$ days for the von Neumann estimator and $289.3$ days vs. $339.1$ days for the Bartels estimator. This smaller value is caused by the presence of the secondary prominent minimum for the FeII line, which is at $\sim 327-335$ days for the von Neumann estimator and at $\sim 327-381$ days for the Bartels estimator, see Figure~\ref{fig_vonneumann_whole} (von Neumann estimator) and Figure~\ref{fig_bartels_whole} (Bartels estimator).   

The analysis performed using the discrete correlation function (DCF) indicates a global peak at $527.5$ days for both the total MgII and FeII emission light curves (specifically for the slot weighting of light curve pairs with the time step of 5 days). When we constructed 400 bootstrap realizations of continuum--line emission pairs, we obtained the peak time delay of $527.0^{+32.0}_{-25.0}$ days for the MgII total emission and $527.0^{+24.0}_{-35.0}$ days for the FeII total emission, i.e. the time delay appears to be the same within uncertainties for both lines. However, the overall time-delay peak distribution is broad with multiple peaks present.  

The time-delay analysis using the $z$-transformed discrete correlation function (zDCF) yields the peak values with large uncertainties, especially for the MgII emission, for which we got $291.3^{+190.2}_{-149.2}$ days in the observer's frame. For the MgII emission, the peak time-delay is actually smaller than the peak time-delay for the FeII emission, for which we obtained $353.6^{+77.6}_{-93.2}$ days. The zDCF value as a function of the time delay is depicted in Figure~\ref{fig_zdcf} for both lines. The zDCF time-delay peak for the MgII line is broader and has a smaller correlation coefficient of $\text{zDCF}=0.60^{+0.16}_{-0.14}$ than the one for the FeII line, $\text{zDCF}=0.68^{+0.13}_{-0.11}$. 

In summary, using different time-delay determination methods, we detect consistently a time-delay peak close to $\sim 520$ days for the MgII emission in the observer's frame, while for the total FeII emission we detect the presence of two time delays, at $\sim 340$ days and $\sim 510$ days, both of which are usually present in the time-delay distributions. The presence of two peaks of comparable height is occasionally noticed in other lines, like H$\beta$ \citep{dupu2015}. When we put together comparable time-delay peaks, we obtain the mean time delay of $523.5^{+23.6}_{-37.1}$ days for the MgII emission and $513.0^{+26.2}_{-48.1}$ and $342.5^{+50.6}_{-57.0}$ days for the FeII emission. When calculated with respect to the rest frame of the source at the redshift of $z=0.90052$, the total MgII-emission time delay is $\tau_{\rm MgII}=275.5^{+12.4}_{-19.5}$ days. For the total FeII-emission, the longer rest-frame time delay is $\tau_{\rm FeII,L}=270.0^{+13.8}_{-25.3}$ days, which is consistent with the MgII time delay within uncertainties. The shorter FeII rest-frame time delay is $\tau_{\rm FeII, S}=180.3^{+26.6}_{-30.0}$ days. The light-travel distance of the MgII emission region is $0.23\,{\rm pc}$. For the FeII region, the set-up is more complex. Its mean distance is comparable to the MgII emission region, $\sim 0.23\,{\rm pc}$, based on the longer time-delay peak. However, the shorter time-delay indicates that the FeII region is more extended in the direction towards the observer, as we will discuss further in Section~\ref{sec_discussion}. Due to the non-zero inclination, a part of the FeII region is located by $(275.5-180.3)c\sim 0.08\,{\rm pc}$ closer to the observer. When illuminated by the same photoionizing radiation, a fraction of the FeII-reprocessed photons reaches the observer sooner than the MgII-reprocessed radiation, while the other part shares the same reprocessing region with MgII.

\subsection{Wavelength-resolved time lags}
\begin{center}
\begin{figure*}
   
    \includegraphics[scale=0.57]{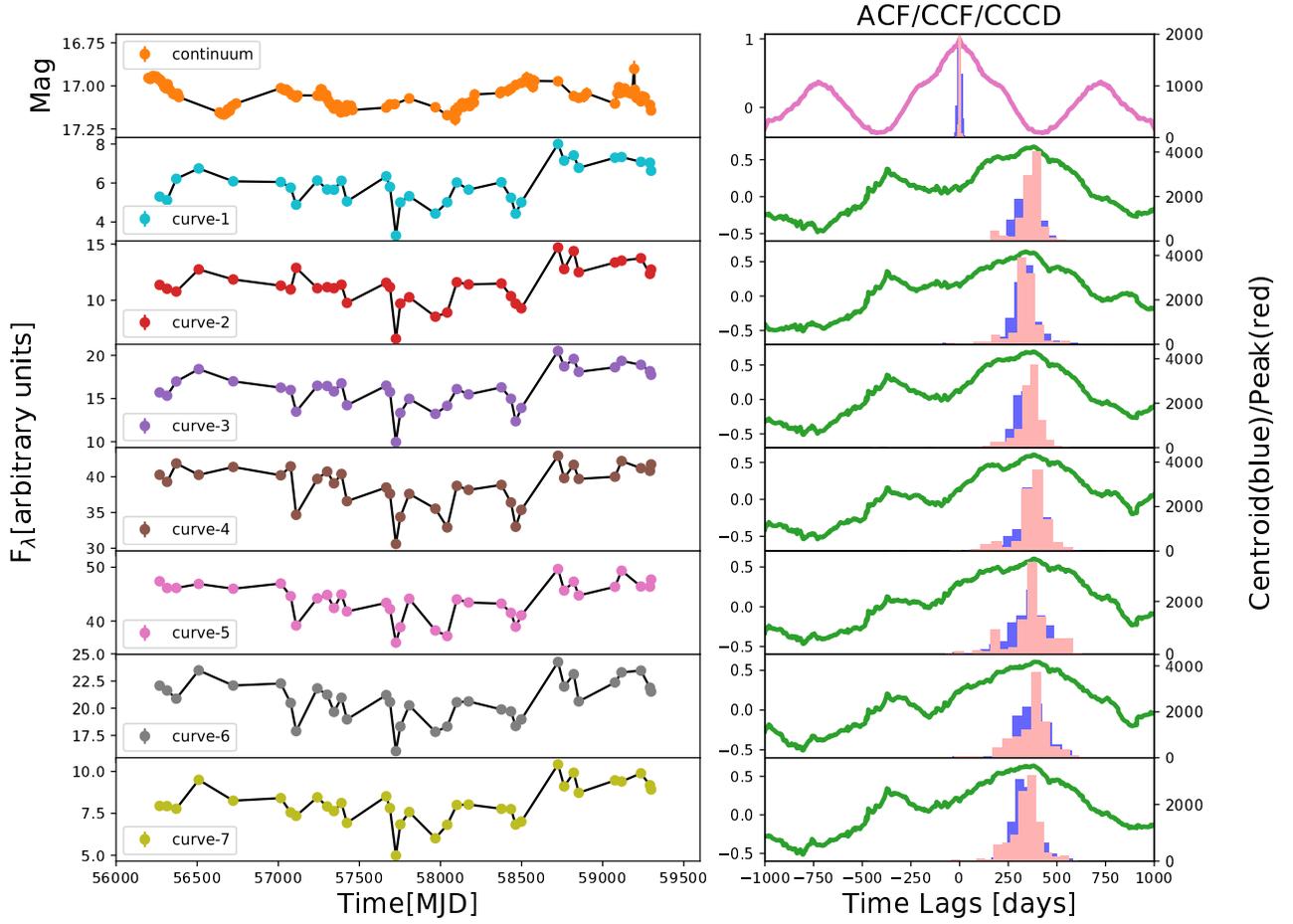}
    \caption{{\tiny {\bf First column:} The top panel shows the continuum light curve in magnitude and the other seven panel show the seven curves in different wavebands derived from the combination of photometric and the spectroscopic observations. The corresponding wavebands range are mentioned in Table \ref{tab:Fvar}. Curve-1 and curve-7 have unit in 10$^{-17}$ and the rest are in 10$^{-16}$. {\bf Second column:} This represents the auto-correlation of continuum and the ICCF results for all the curves with respect to continuum. Histograms are the peak (red) and centroid (blue) distribution from the ICCF with 10000 bootstrap realizations.}}
    \label{fig:lightcurves}
\end{figure*}
\end{center}

As discussed in Section 3.3, seven light curves are created from the different part of the RMS spectrum.
The lighcurves are shown in Figure~\ref{fig:lightcurves}. The original curves are properly normalized, but for a better comparison with photometry they are also plotted in magnitude scale, with arbitrary normalization.

To understand the kinematics and the geometry of broad line region (BLR) in quasars, velocity-resolved time lags are essential. Here we present our investigation of BLR kinematics in a bright quasar CTS C30.10 using its MgII+FeII emission, with the continuum power-law subtracted. The MgII+FeII combination is divided into seven different velocity bins after subtraction of the continuum power-law. The corresponding seven light curves have been produced and eventually the time delay with respect to the $V$-band continuum light curve was investigated using different methodologies. The methods described in Subsection~\ref{subsec_timelag_methods} were used to estimate the time lags (and the uncertainty) between the various curves and the continuum. The results are summarized in Table~\ref{tab:curves1_7}.\\

\begin{figure*}
    \centering
    \includegraphics[width=\columnwidth]{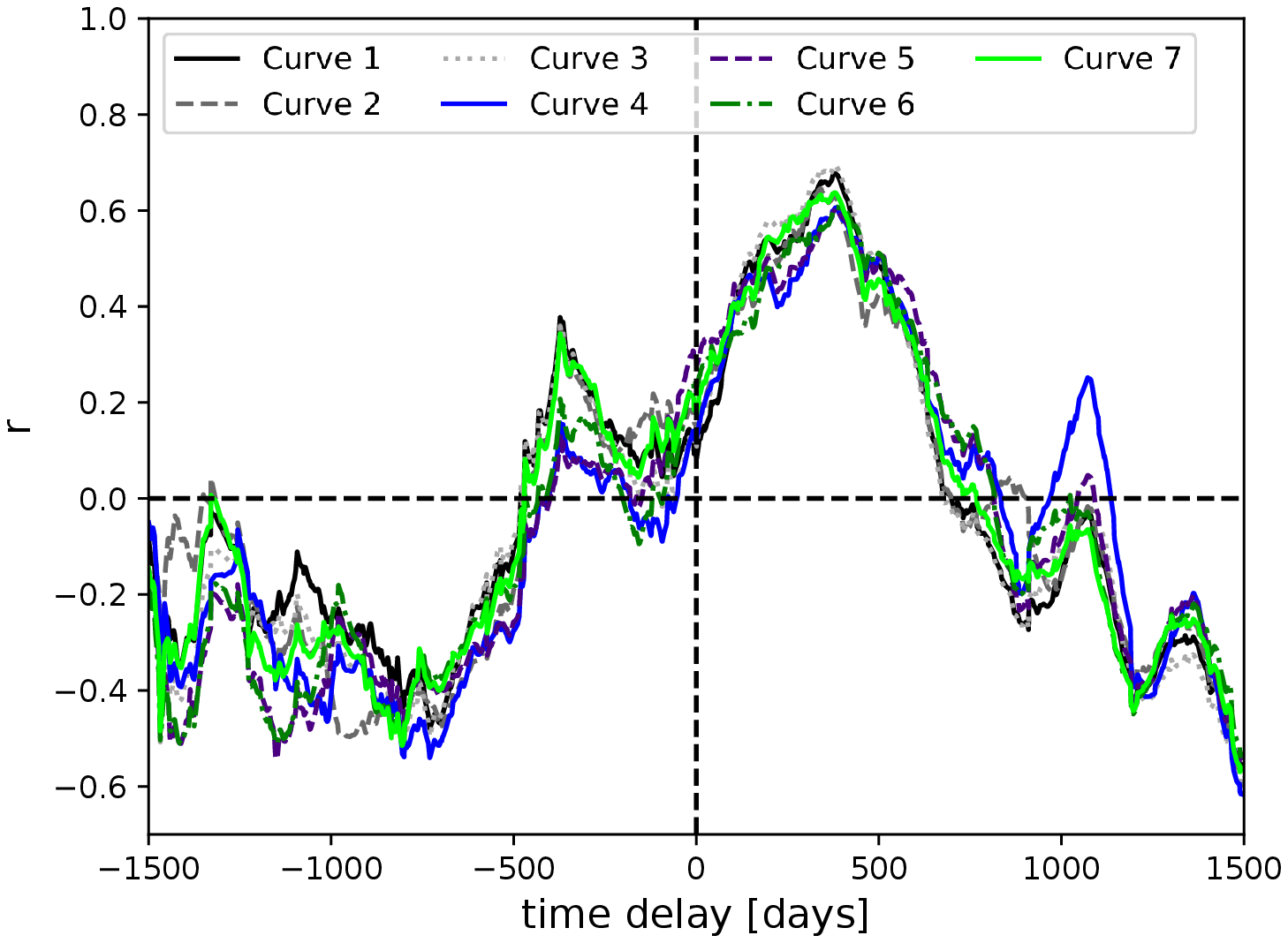}
    \includegraphics[width=\columnwidth]{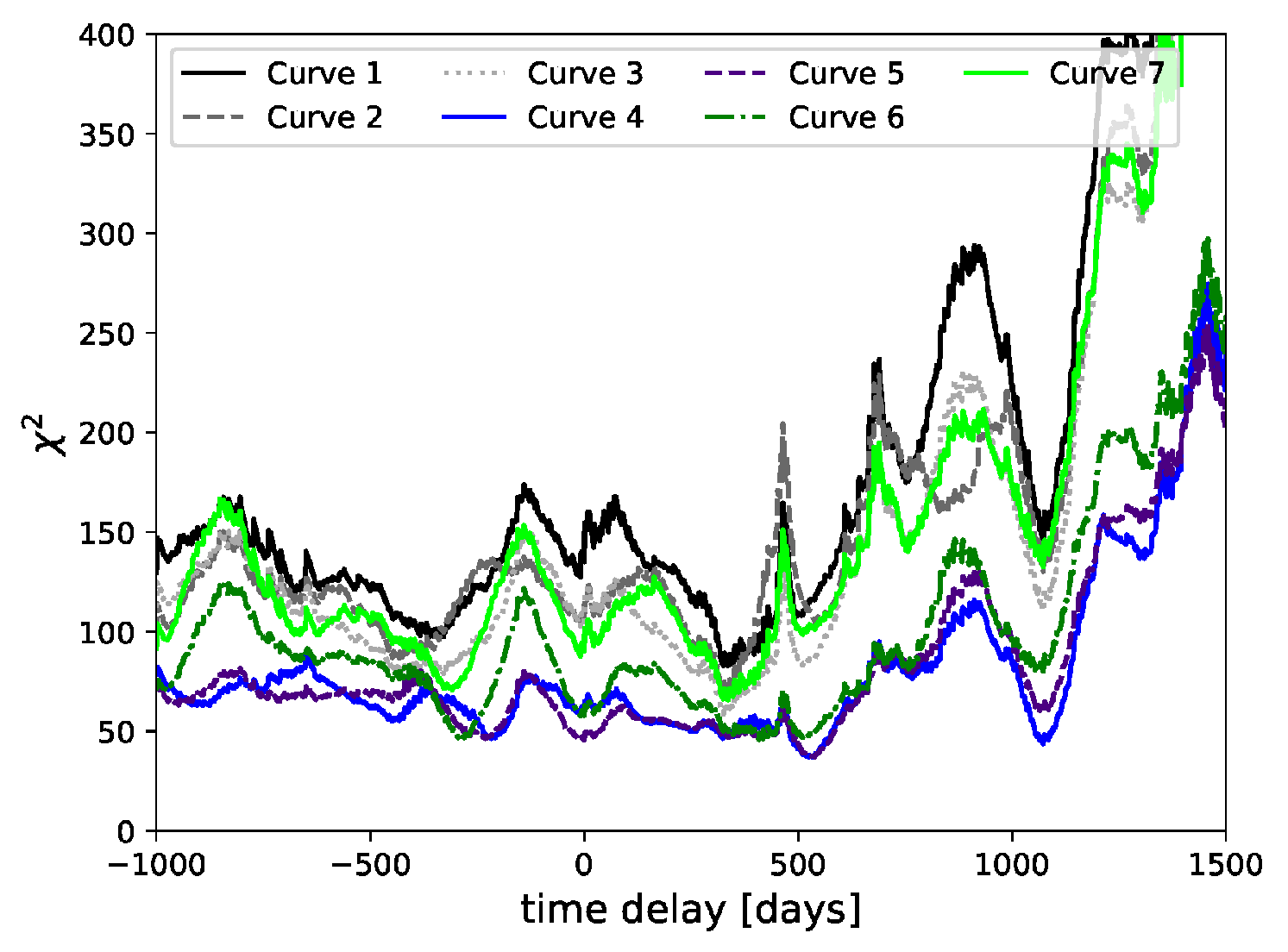}
    \caption{Interpolated Cross-Correlation Function (ICCF) and $\chi^2$ values as a function of time delay. {\bf Left panel:} The ICCF as a function of time delay (expressed in days in the observer's frame) for the seven light curves according to the legend. {\bf Right panel:} The $\chi^2$ value as a function of the time delay (in days) in the observer's frame for the seven light curves according to the legend.}
    \label{fig_iccf_chi2}
\end{figure*}

{\bf ICCF:} The Interpolated Cross-Correlation function (ICCF) is shown for each light curve in Figure~\ref{fig_iccf_chi2} (left panel). According to the time delay values inferred from the maximum correlation coefficient, see Table~\ref{tab:curves1_7}, the time delay is between $340$ and $380$ days in the observer's frame, with the weak increase for light curves 4, 5, and 6. The maximum correlation coefficient is $\sim 0.6-0.7$. The ICCF uses the flux and amplitude randomization technique to estimate the time lag distribution, in particular its centroid and the peak. We estimated the centroid and the peak time lag and corresponding plots are shown in Figure \ref{fig:lightcurves}. For all the curves the time lags from the centroid are consistent within the errorbar between $\sim$330--370 days. However, the time lags from the peak distribution is higher in all the curves compared to centroid values and it lies between $\sim$340 -- 385 days. In both the cases, the longer time lags are noted in curve 4, 5 $\&$ 6 which is expected as they represents the MgII part of the spectrum.\\

$\mathbf{\chi^2}$: This method is based on the $\chi^2$-minimization technique. The results recovered from this method are generally consistent with the ICCF results. The time delays were found to be between $\sim$324 -- 538 days with higher time delays for light curves 4, 5, $\&$ 6. The $\chi^2$ values as a function of time delay for individual light curves are shown in Figure~\ref{fig_iccf_chi2} (right panel). The higher time delays are consistent with the RMS spectrum shown in Figure~\ref{fig:t-delay} where the curve-4 and curve-5 lie exactly in the middle of MgII line emission. The bootstrap technique is also applied to obtain the time-delay peak distribution. We generate 10000 light curve pairs based on the actual seven light curves. Based on the time-delay peak distribution, we determine the final peak and the mean of the distribution. The final peak asymmetric errorbars are inferred from the left and the right standard deviations within $30\%$ of the main peak surroundings. The mean time-delay follows a similar trend of increasing and decreasing time delays towards longer wavelengths, with the largest time delay of $558.3$ days for light curve 4.\\

{\bf Data regularity estimators (von Neumann, Bartels):} When we apply the data regularity estimators (von Neumann, Bartels) to the seven light curves, we obtain the minimum estimator value at $\sim 512$ days in the observer's frame both for von Neumann and Bartels estimators, see Table~\ref{tab:curves1_7} and Figure~\ref{fig_vonneumann_bartels}. However, the estimator profile as a function of the time delay changes qualitatively close to this global minimum -- closer to the MgII line wings, the broad minimum is shallower, see Figure~\ref{fig_vonneumann_bartels} for the von-Neumann estimator (left panel) and the Bartels estimator (right panel), which also results in the smaller mean value of the peak time delay as inferred from the bootstrap analysis (1000 realizations). This is in contrast to light curves 3, 4, and 5 close to the line center, where the minimum at $\sim 512$ days is more pronounced, resulting in the larger mean time-delay values.\\

{\bf DCF:} Investigating the time lags using the DCF method yielded shorter time lags for all the seven light curves in comparison with other methods. The dominant peak in the observer's frame is at $187.5$ days both for the default DCF investigation using the observed light curves as well as in the peak distribution inferred from 200 bootstrap realizations for each light curve. In Table~\ref{tab:curves1_7}, we separately list the peak values for the DCF evaluation using the slot- and the Gauss-weighting of the light curve pairs, where the time bin is constant and we set it to 25 days. For the bootstrap runs, we separately calculate the peak and the mean values of the corresponding time-lag peak distributions. While the peak is always close to $187.5$ days, the mean value shifts towards larger values for light curves 4 and 5 since the secondary time-lag peak at $\sim 450-550$ days becomes more prominent for these light curves. \\

{\bf zDCF:} For the zDCF method, the measured time-delays are in agreement within the uncertainties. The peak values of the time-delay are $327.9$ days in the observer's frame for the first three wavebands, then they increase to $370.5$ days for the band 4, 5, and 6. The emission light curve 7 corresponding to the red wing of the line has the peak time delay again at $327.9$ days. The zDCF values as a function of the time delay in the observer's frame are depicted in Figure~\ref{fig_zDCF_all} for individual light curves and results are presented in Table~\ref{tab:curves1_7}. The shift of $42.6$ days between the time-delay peaks of the first (as well as the second, third, and the seventh light curves) and the fifth light curve (as well as the fourth and the sixth light curves) is highlighted by the corresponding horizontal lines.\\

{\bf JAVELIN:} The results from the JAVELIN is 
shown in Figure \ref{fig:javelin} and recovered time delays are much higher than the ICCF. For most of the curves (1, 3, 4, 5, 6 $\&$ 7) the recovered time delays are between $\sim$502 -- 529 days. For curve 2, we found rather a small time delay of the order of 195 days, much smaller than ICCF also. 
To quantify the errorbars on the time delay results, we applied the bootstrap technique for 1000 realizations and estimated the peak time delay with one-sigma errorbar. 
The results are presented in Table~\ref{tab:curves1_7}. Time delays corresponding to various curves are shown with RMS spectrum in Figure \ref{fig:t-delay}, and it agree with $\chi^2$ methodology for most of the curves.
\\

\subsection{Summary of the wavelength-dependent trends and BLR kinematics}

\begin{figure}
    \centering
    \includegraphics[scale=0.46]{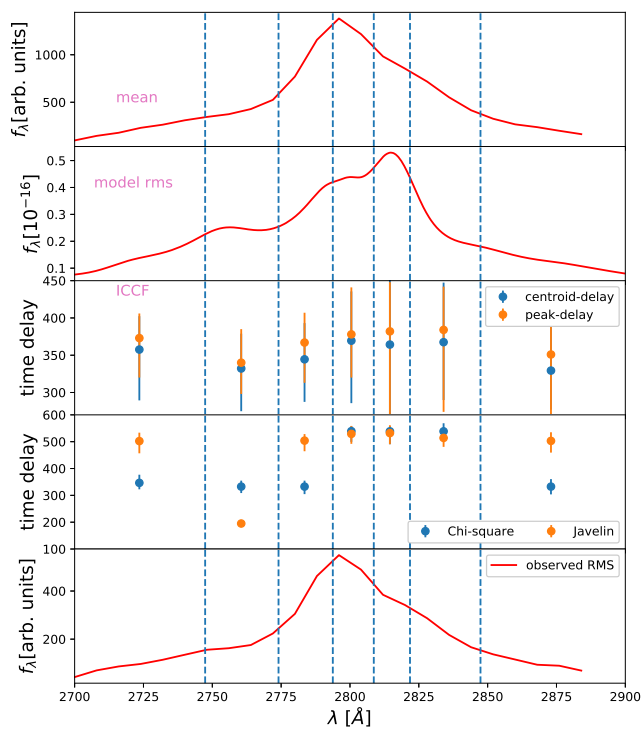}
    \caption{Here we show the wavelength dependent time lags from various methods along with the mean and RMS spectrum. Time lags seem to weakly follow the RMS spectrum. }
    \label{fig:t-delay}
\end{figure}

\begin{figure*}
    \centering
     \includegraphics[width=\columnwidth]{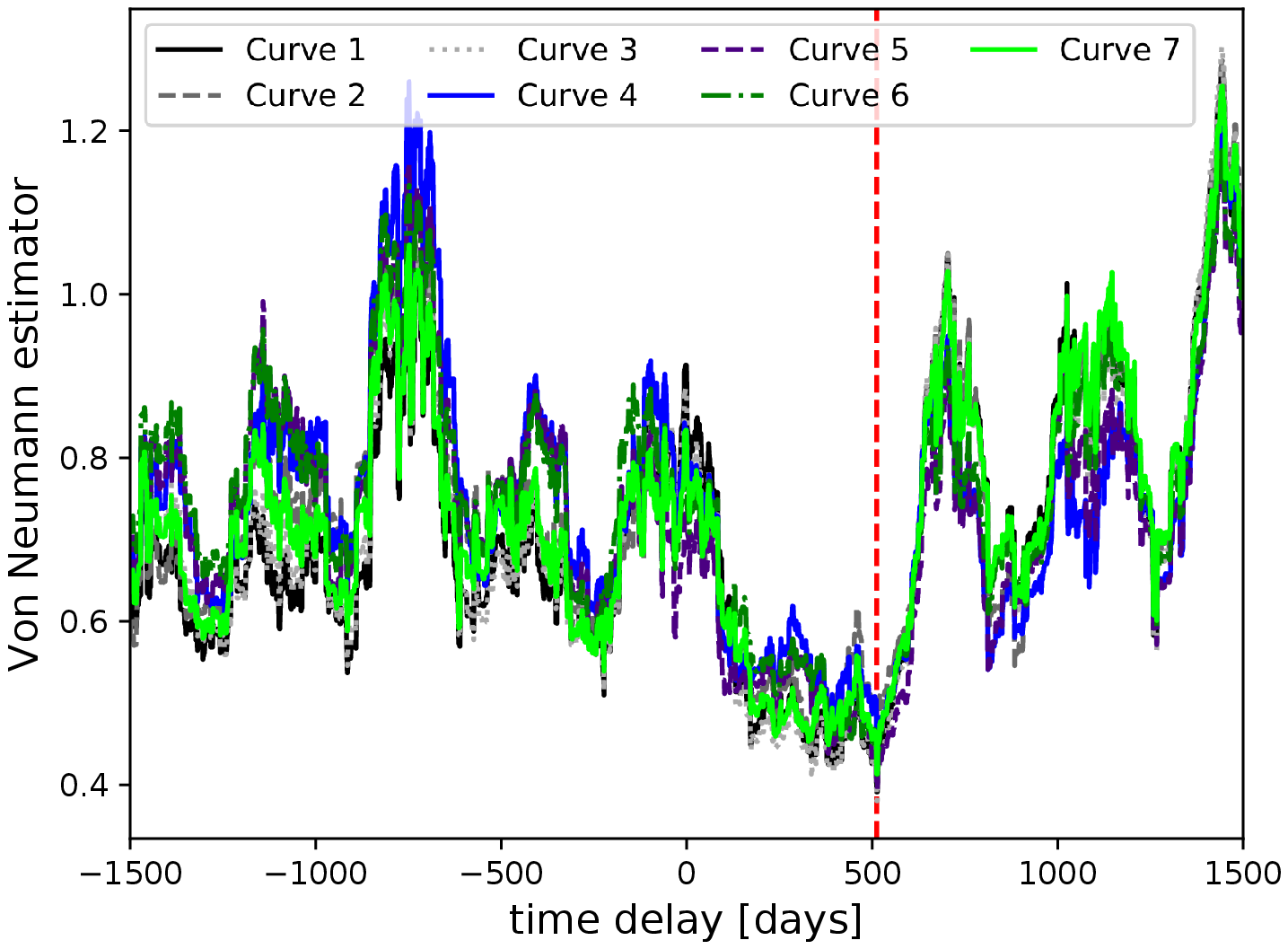}
    \includegraphics[width=\columnwidth]{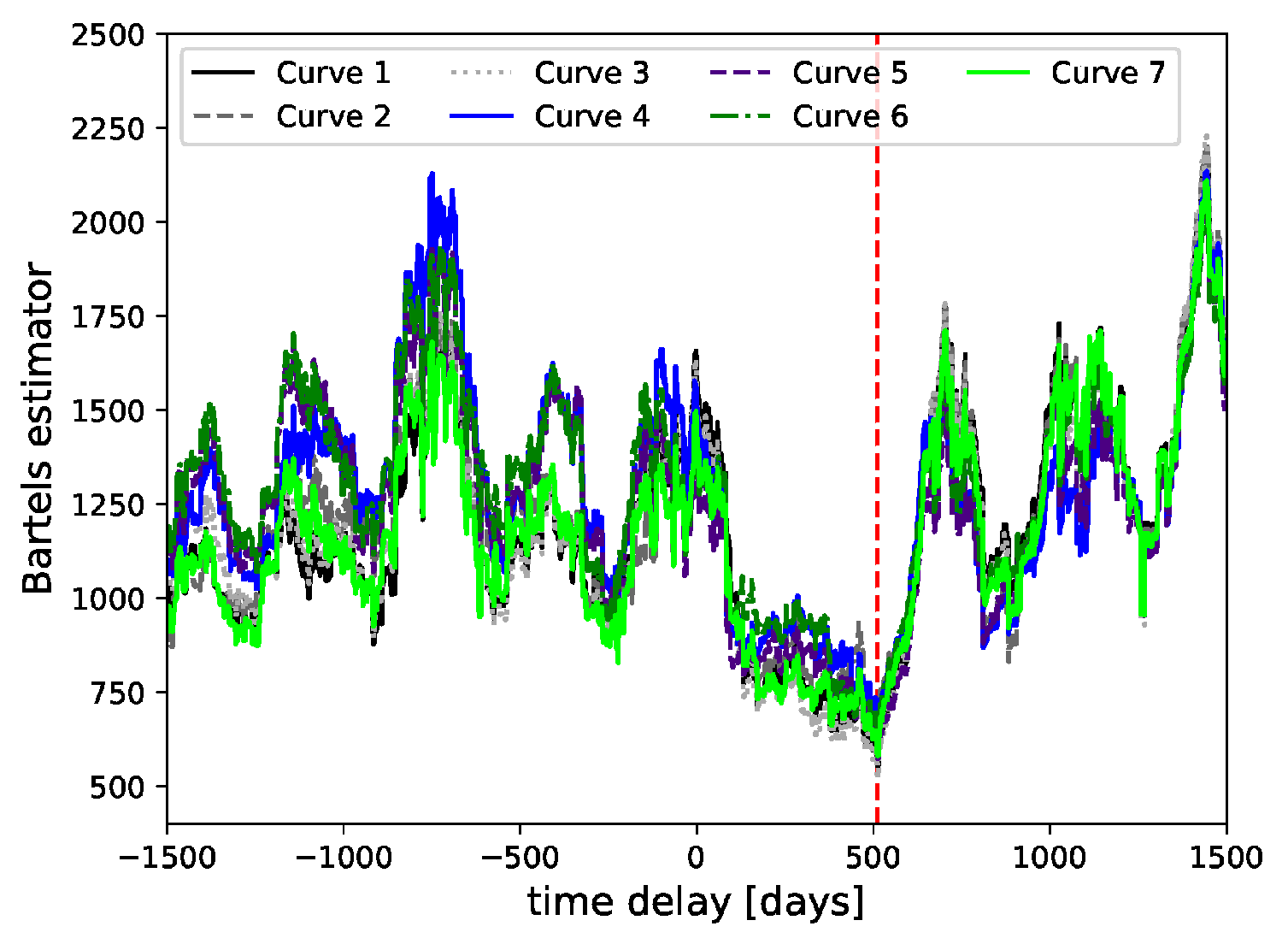}
    \caption{Estimators of data randomness/regularity as a function of the time delay in the observer's frame. {\bf Left panel:} Von Neumann estimator for seven MgII line light curves according to the legend. {\bf Right panel:} The Bartels estimator value for the same MgII light curves in the observer's frame. The red dashed line represent the estimated time delay.}
    \label{fig_vonneumann_bartels}
\end{figure*}

\begin{figure}
    \centering
    \includegraphics[width=\columnwidth]{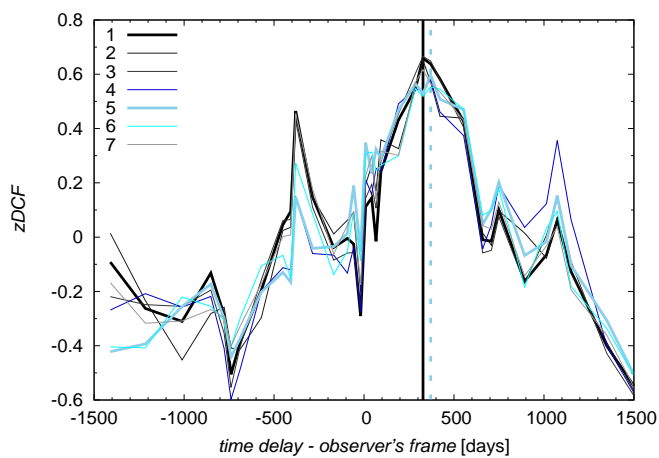}
    \caption{The $z$-transformed DCF as a function of the time delay in the observer's frame for individual light curves according to the legend. Overall, there is a time-delay peak between 300 and 400 days consistent within uncertainties for all the light curves. We highlight the shift of $\sim 40$ days between the time-delay peaks of the first and the fifth waveband by using the corresponding horizontal lines to denote the peak values.}
    \label{fig_zDCF_all}
\end{figure}

\begin{figure*}
\centering
\includegraphics[scale=0.35]{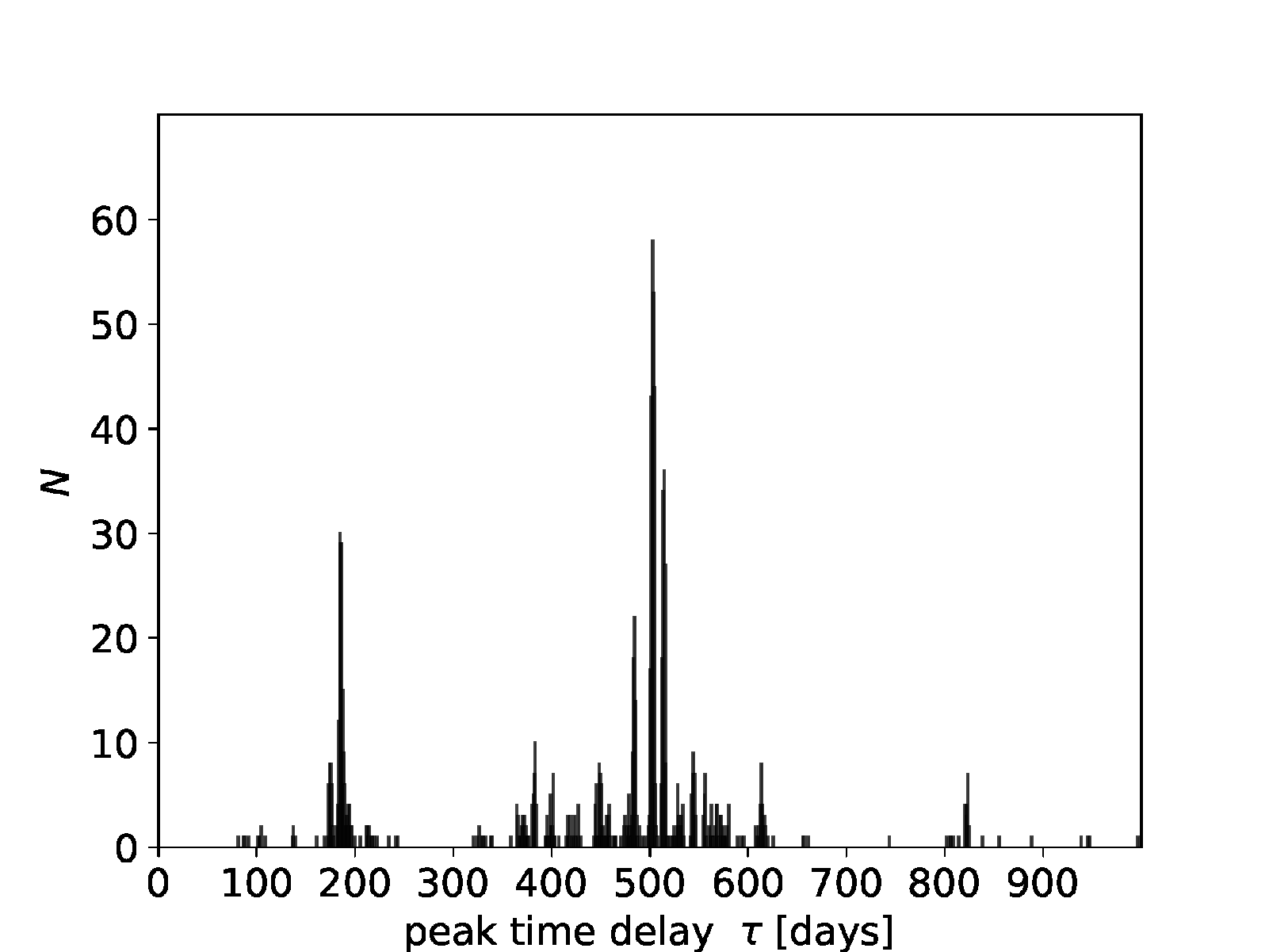}
\includegraphics[scale=0.35]{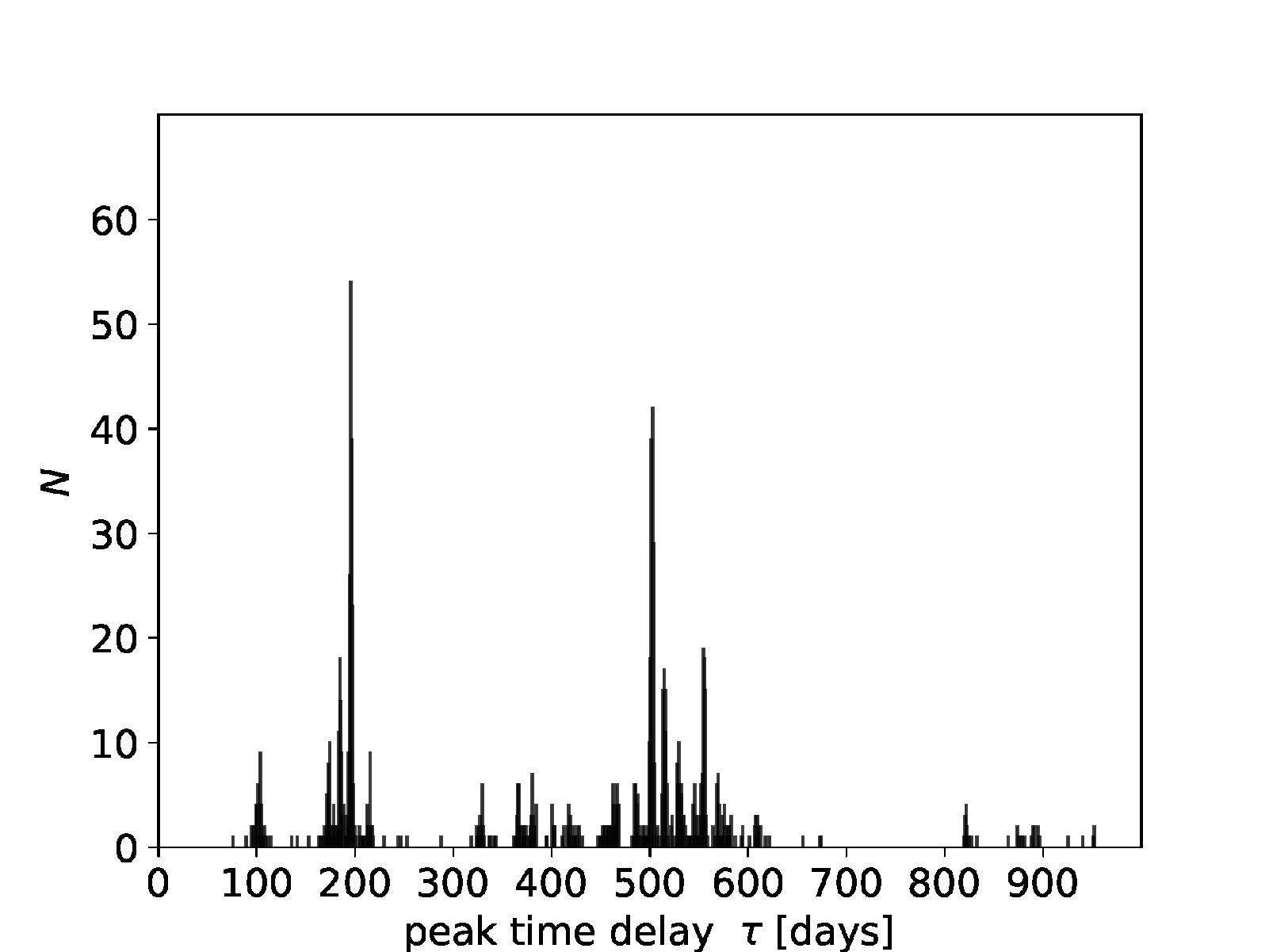}
\includegraphics[scale=0.35]{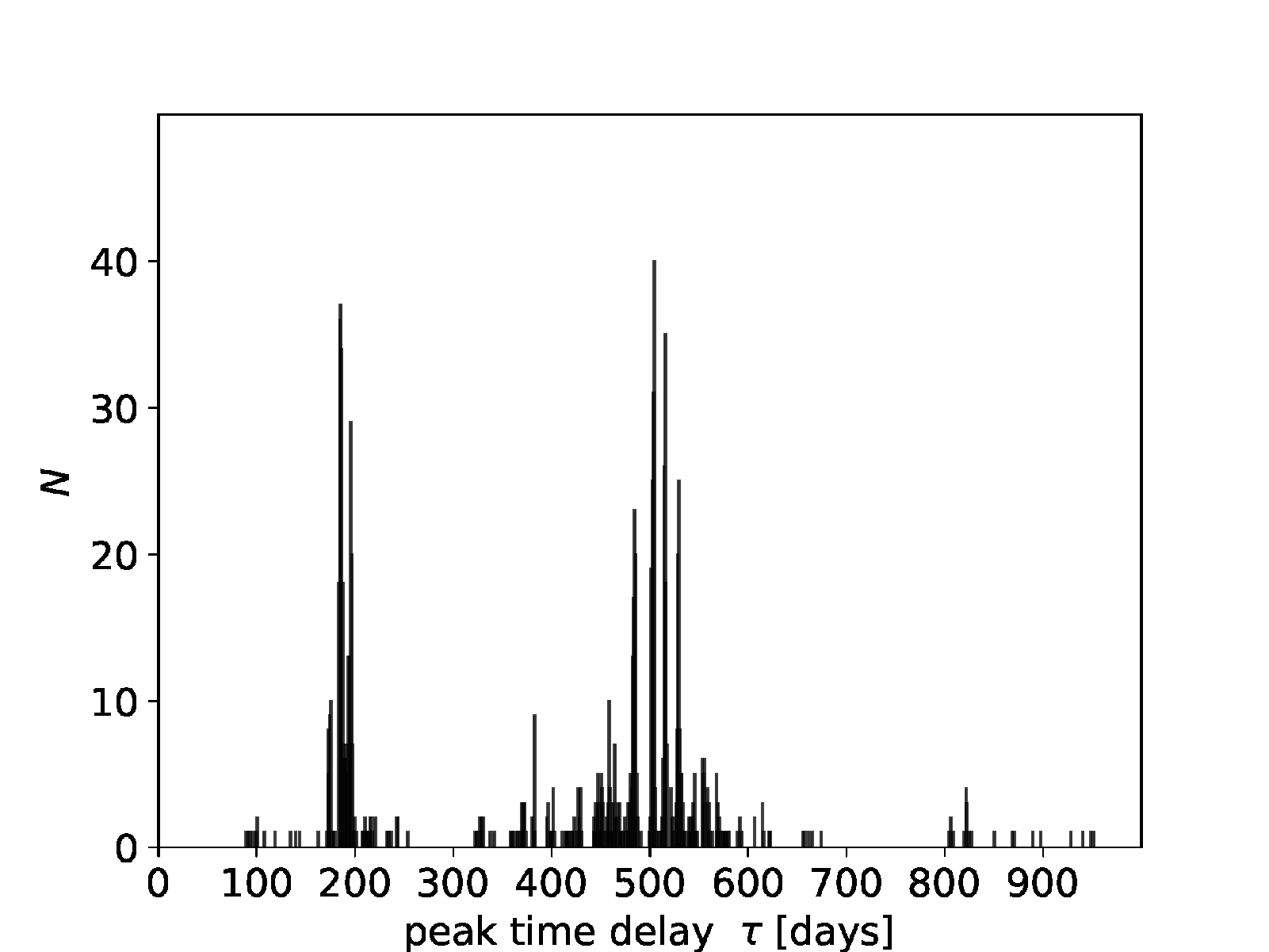}
\includegraphics[scale=0.35]{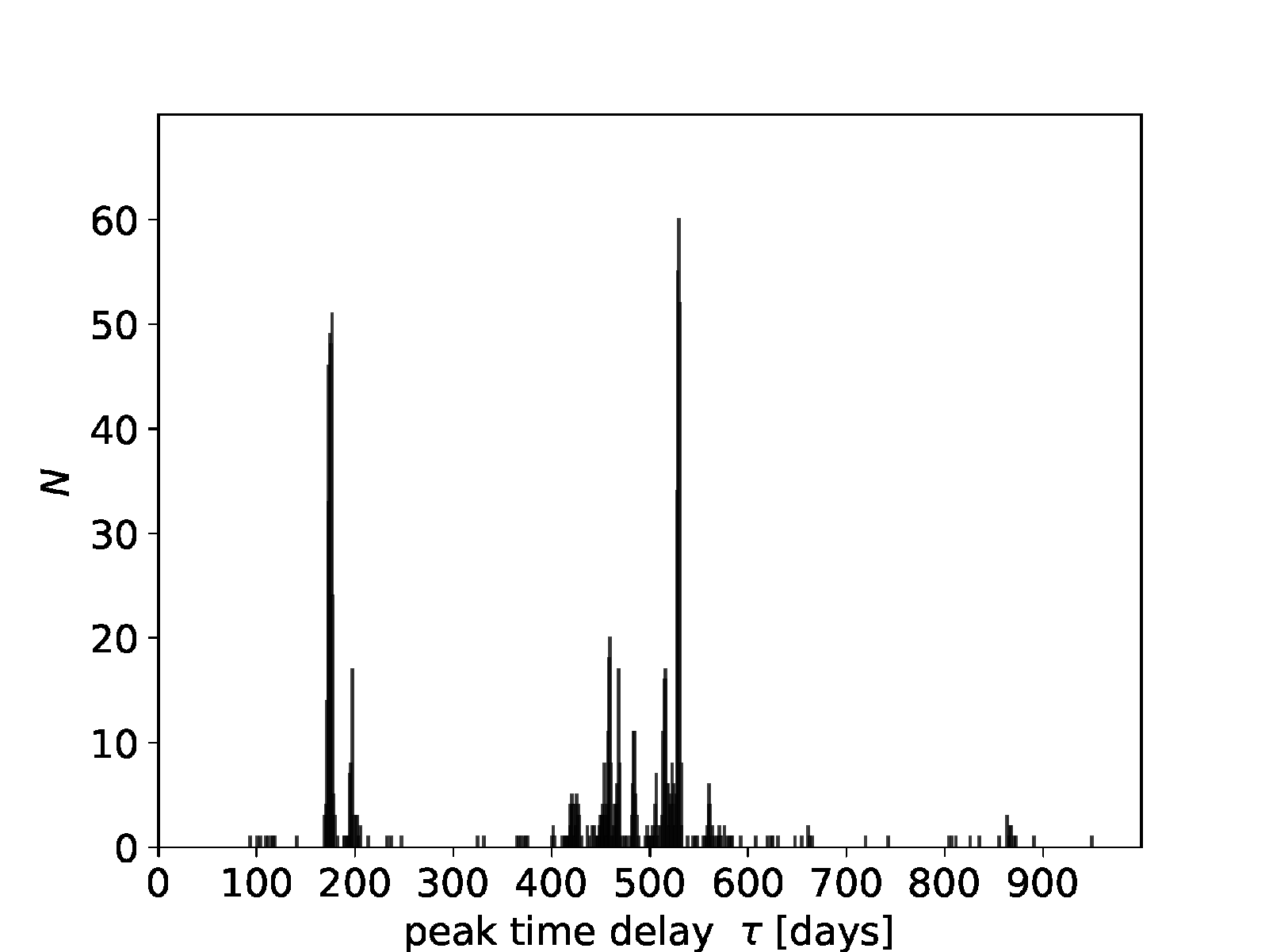}
\includegraphics[scale=0.35]{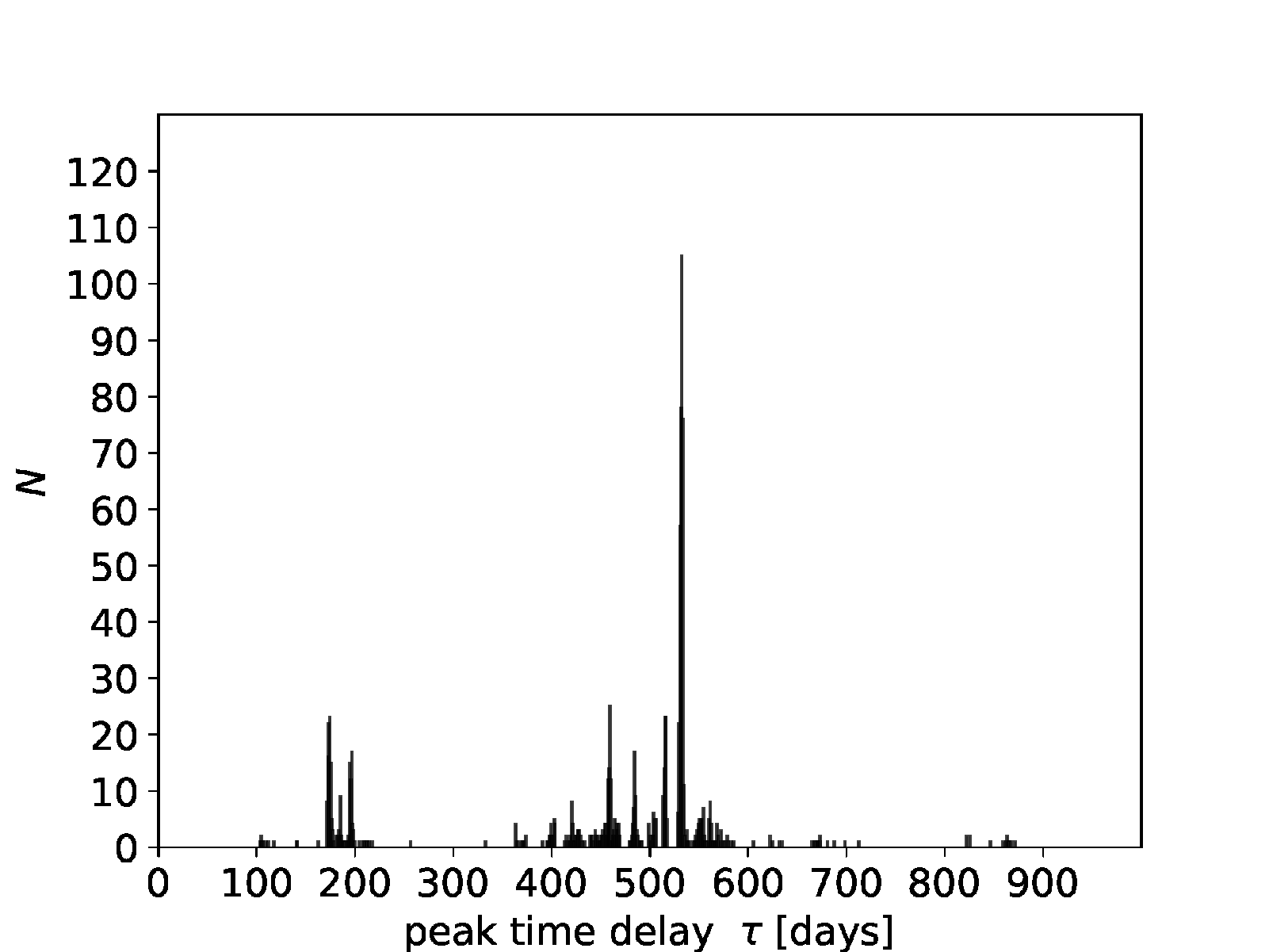}
\includegraphics[scale=0.35]{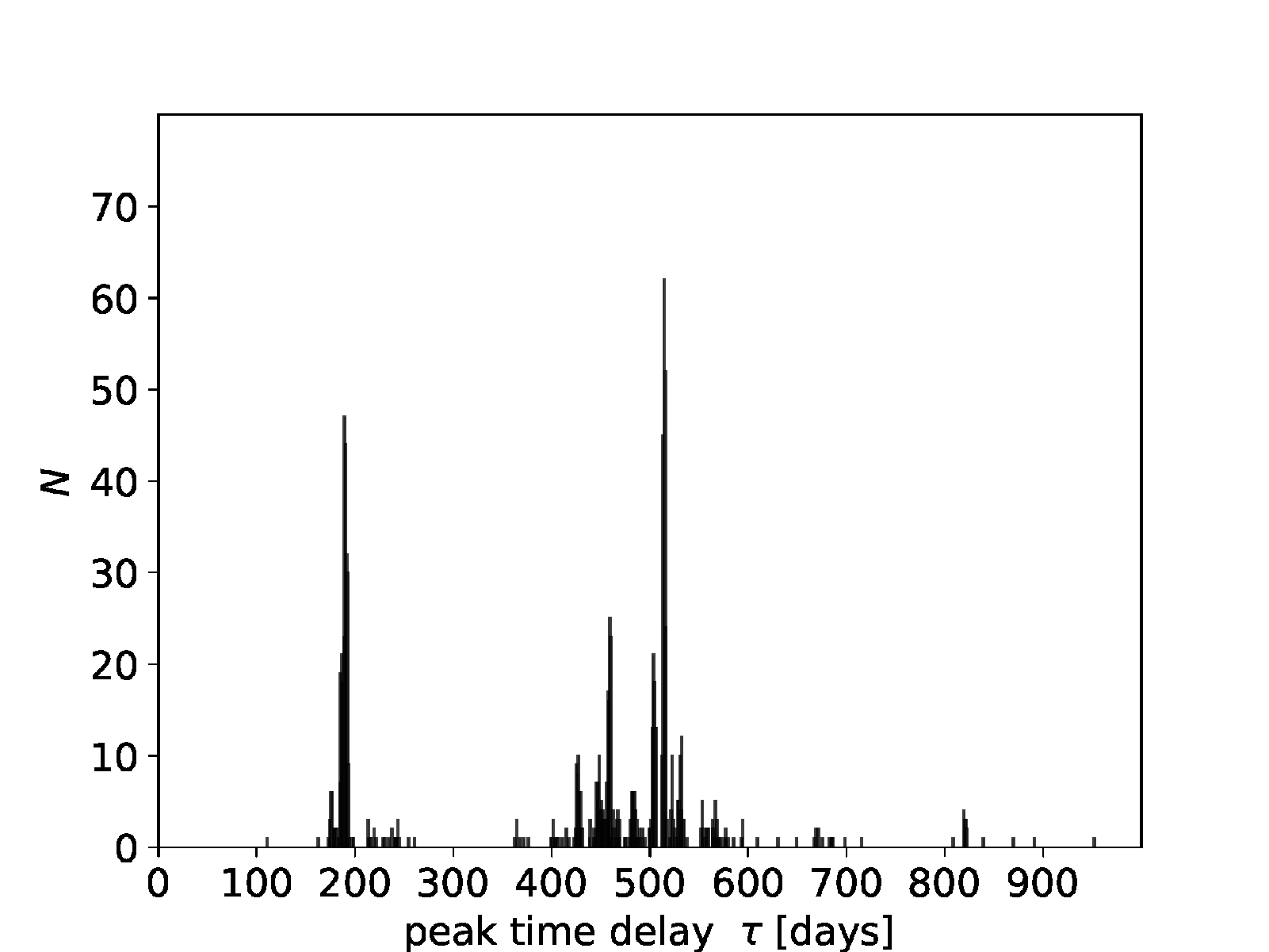}
\includegraphics[scale=0.35]{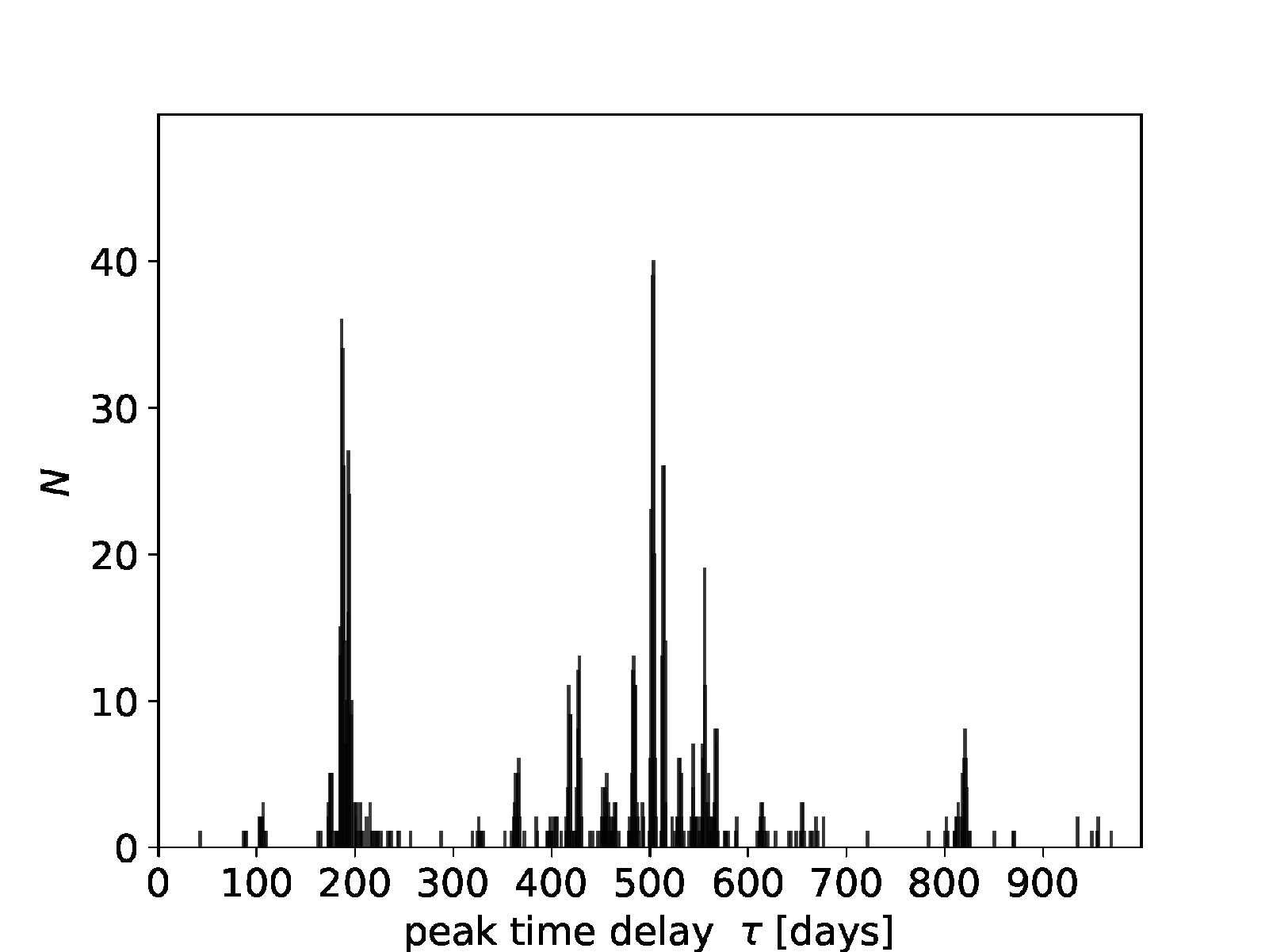}
\caption{Javelin bootstrap results with 1000 realizations for all the seven curves (from left to right). The peak and  results from this is mentioned in Table 4.}
\label{fig:javelin}
\end{figure*}

The visual summary of the observed trends is given in Figure~\ref{fig:t-delay}. The time delays in the curves dominated by the MgII emission are longer than for FeII-dominated curves, specifically 1 or 7. The two-component character of MgII line does not show up. Component 1 dominated the curve 4 while component 2 dominates curve 5, but we do not see any differences between the time delays for these two curves. This supports the conclusion that the need for two components rather reflects a more complex line shape than the actual existence of the two physically separated regions, as was already argued by \citet{Modzelewska2014} at the basis of the mass measurement consistency. Some level of asymmetry of the MgII line is frequently seen, and the two-component fits are required \citep{marziani2013}, but these authors show that nevertheless MgII is a better virial indicator of the BLR motion than H$\beta$ since the centroid shifts with respect to rest frame are lower.

In our wavelength-resolved data we do not see any clear anisotropy along the line shape, the time delay neither decreases or increases systematically with the wavelength.
This means that we do not detect traces of an inflow or an outflow. We did not attempt to study the MgII shape separately since subtraction of FeII would anyway lead to large errors in the line wings so we did not expect any new results from such an approach with the current dataset.

\begin{table*}[]
    \centering
    \caption{Time delay measurements for seven wavelength bins containing a combination of MgII and FeII emission, after subtraction of the power law component. Curve 1 and 7 contain only FeII, curves 4 and 5 are strongly dominated by MgII.}
\resizebox{\textwidth}{!}{      
    \begin{tabular}{c|ccccccc}
\noalign{\smallskip}
\hline\noalign{\smallskip}
with Obs flux & Curve 1 & Curve 2 & Curve 3 & Curve 4& Curve 5 & Curve 6 & Curve 7 \\
\noalign{\smallskip}
\hline \noalign{\smallskip}
ICCF (max $r$) & 381.0 (0.68) & 340.0 (0.65)   & 380.0 (0.69)    & 382.0 (0.61)  & 384.0 (0.61)    & 384.0 (0.61)   & 379.0 (0.64)   \\
\noalign{\smallskip}
ICCF (Centroid) & 357.6$^{+44.8}_{-68.2}$&332.2$^{+46.7}_{-57.3}$&344.5$^{+48.6}_{-57.1}$&369.5$^{+66.1}_{-83.8}$&364.3$^{+93.6}_{-102.3}$&367.5$^{+79.9}_{-77.6}$&329.4$^{+64.1}_{52.4}$\\
\\
ICCF (peak)& 373.0$^{+33.0}_{-53.0}$&340.0$^{+45.0}_{-42.0}$&367.0$^{+40.0}_{-54.0}$&378.0$^{+63.0}_{-58.0}$&382.0$^{+66.0}_{-124.0}$&384.0$^{+58.0}_{-110.0}$&351.0$^{+56.0}_{-88.0}$ \\
\noalign{\smallskip}
\hline \noalign{\smallskip}
Javelin (minimum 1 run) &504.0$^{+4.7}_{-3.2}$ &504.0$^{+13.9}_{-0.5}$ &504.0$^{+13.4}_{-0.4}$&530.0$^{+0.8}_{-5.8}$&533.0$^{+4.2}_{-3.9}$ & 505.0$^{+10.4}_{-0.5}$ & 504.0$^{+12.6}_{-0.4}$\\
\noalign{\smallskip}
Javelin (bootstrap, 1000 run, peak) &502.0$^{+31.3}_{-45.6}$&195.0$^{+9.0}_{-8.0}$ &504.0$^{+24.3}_{-39.6}$ &529.0$^{+29.7}_{-37.0}$ &532.0$^{+29.2}_{-41.3}$ & 514.0$^{+24.0}_{-33.3}$ & 503.0$^{+32.1}_{-43.7}$ \\
\noalign{\smallskip}
\noalign{\smallskip}
\hline \noalign{\smallskip}
{\bf $\chi^2$} (minimum for 1 run) &  333.5 & 324.5 & 324.5  & 524.6  &  538.6 & 407.6 & 336.5    \\
$\chi^2$ - peak (bootstrap) & $346.5^{+30.4}_{-23.9}$  & $332.6^{+22.2}_{-24.0}$  & $332.6^{+28.3}_{-28.2}$  & $539.3^{+18.5}_{-40.1}$  & $538.6^{+19.3}_{-48.1}$  & $538.6^{+30.1}_{-36.8}$ & $332.6^{+27.9}_{-29.0}$ \\
\noalign{\smallskip}
$\chi^2$ - mean (bootstrap) & $371.2^{+229.0}_{-118.9}$  & $355.3^{+401.4}_{-133.6}$  & $374.3^{+286.2}_{-99.2}$  & $558.3^{+270.7}_{-159.4}$  & $439.9^{+313.3}_{-147.2}$  & $418.5^{+266.7}_{-121.2}$ & $347.4^{+250.8}_{-135.7}$\\
\noalign{\smallskip}\\
\hline \noalign{\smallskip}
von Neumann  (minimum for 1 run) & 512.0 & 335.0   & 512.0    & 512.0  & 512.0    & 512.0    & 512.0    \\
von Neumann -- peak (bootstrap) & $511.3^{+15.6}_{-47.5}$ & $511.3^{+26.0}_{-50.7}$  & $511.3^{+12.1}_{-46.7}$  & $511.3^{+24.0}_{-38.1}$  & $511.3^{+21.9}_{-45.3}$  & $511.3^{+10.3}_{-44.8}$  & $511.3^{+13.7}_{-50.3}$\\
\\
von Neumann -- mean (bootstrap) & $357.4$ & $286.4$   & $393.3$  & $351.9$  & $347.6$  & $363.4$   & $312.7$\\
\noalign{\smallskip}
\hline \noalign{\smallskip}
\noalign{\smallskip}
Bartels  (minimum for 1 run) & 512.0 & 512.0   & 512.0    & 512.0  & 512.0  & 512.0   & 512.0    \\
\noalign{\smallskip}
Bartels - peak (bootstrap) & $511.9^{+15.5}_{-46.5}$ &  $511.8^{+30.0}_{-51.7}$     &  $511.2^{+13.2}_{-47.2}$      &  $511.9^{+24.6}_{-42.1}$      & $511.8^{+22.2}_{-43.8}$       & $511.7^{+16.2}_{-39.7}$     & $511.7^{+20.0}_{-50.3}$   \\
\noalign{\smallskip}
Bartels - mean (bootstrap) & $343.5$     & $293.1$      &   $368.2$     & $368.9$       &  $402.1$      & $358.8$     & $321.8$   \\
\hline\noalign{}\\
DCF (25 day; slot) &   187.5  &   187.5    &   187.5   &   187.5   &  187.5  & 187.5 & 187.5 \\
DCF (25 day; gauss) &  187.5  &   312.5    &   187.5   &   187.5   &  187.5  & 187.5 & 212.5 \\
DCF (25 day; slot; bootstrap-peak) & $187.5^{+26.7}_{-37.5}$ & $187.5^{+26.7}_{-0.0}$       & $187.5^{+41.0}_{-0.0}$       &  $187.5^{+19.5}_{-22.4}$    & $187.5^{+30.5}_{-25.0}$     & $187.5^{+11.3}_{-0.0}$    & $187.5^{+0.0}_{-0.0}$    \\
\noalign{\smallskip}
DCF (25 day; slot; bootstrap-mean) & $214.9$ & $113.4$       & $117.1$       &  $150.1$    &  $131.0$    & $37.5$    & $122.6$   \\
\noalign{\smallskip}
DCF (25 day; gauss; bootstrap-peak) & $187.5^{+37.2}_{-23.1}$ & $212.5^{+30.8}_{-23.4}$      & $187.5^{+36.8}_{-23.9}$       & $187.5^{+40.0}_{-18.5}$     & $187.5^{+35.4}_{-5.6}$     & $187.5^{+23.8}_{-19.3}$    & $212.5^{+25.4}_{-13.7}$    \\
\noalign{\smallskip}
DCF (25 day; gauss; bootstrap-mean) & $165.0$ & $150.9$      & $133.9$       & $265.6$     & $204.5$     & $111.3$    & $65.9$    \\
\hline\noalign{\smallskip}\\
zDCF (min. 20 points per bin) & $327.9^{+109.9}_{-74.6}$ &  $327.9^{+116.9}_{-71.6}$   & $327.9^{+99.0}_{-76.3}$ & $370.5^{+114.8}_{-139.2}$    &  $370.5^{+139.7}_{-132.1}$  &  $370.5^{+137.4}_{-109.4}$  & $327.9^{+132.8}_{-72.8}$   \\
\noalign{\smallskip}
\hline
\hline

    \end{tabular}
}    
    \label{tab:curves1_7}
\end{table*}



\begin{figure}
    \centering
    \includegraphics[scale=0.5]{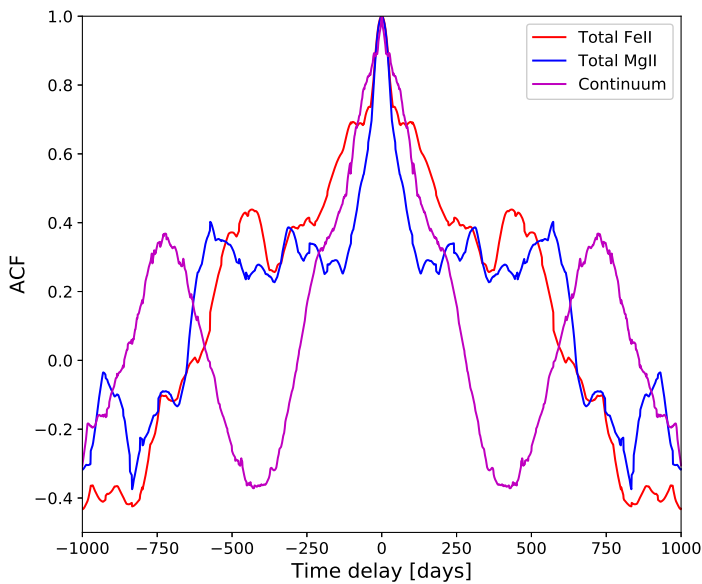}
    \caption{Auto-correlation function of total MgII and total FeII along with continuum.}
    \label{fig:ACF}
\end{figure}

\begin{figure*}
    \centering
    \includegraphics[scale=0.5, angle=-90]{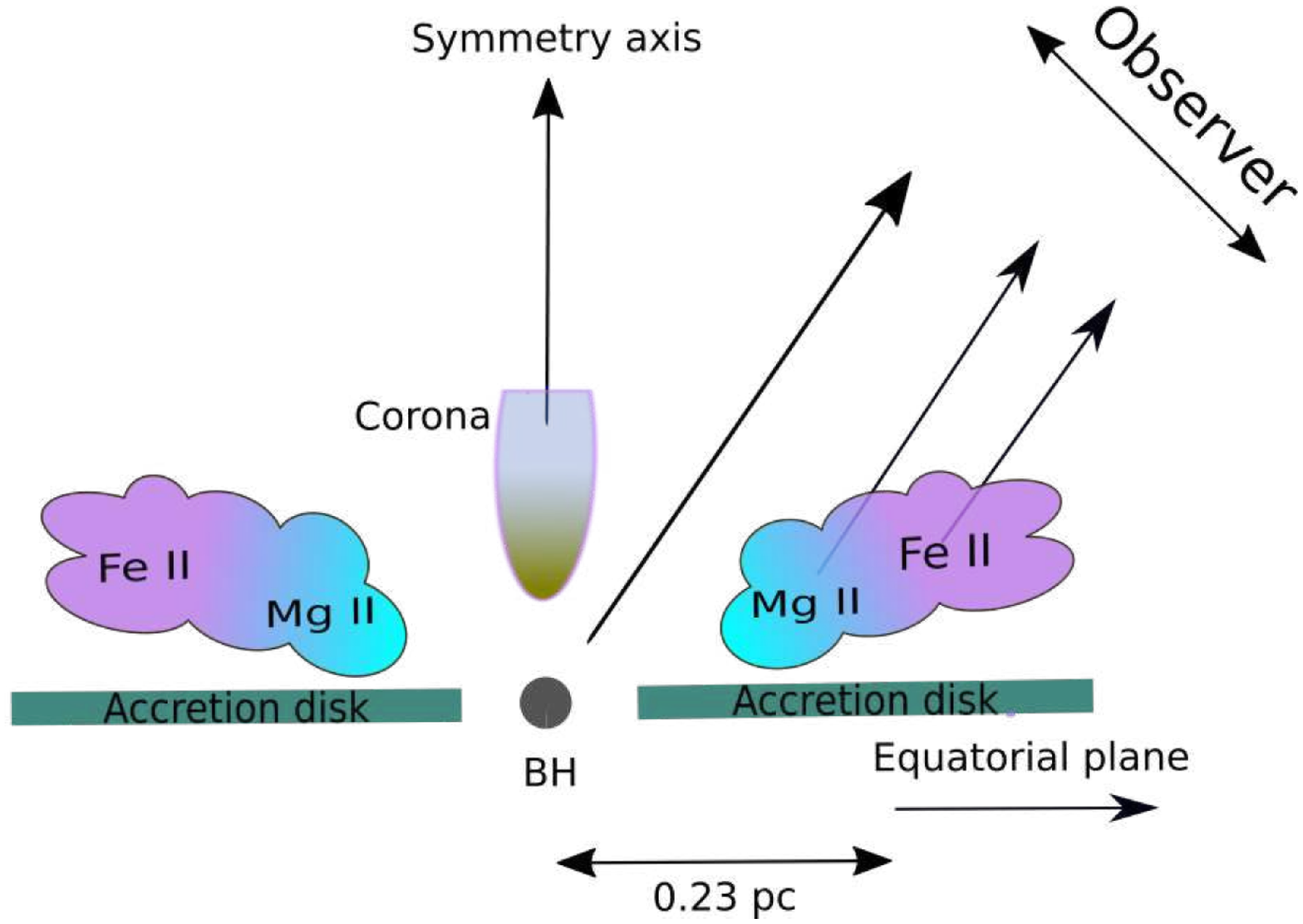}
    \caption{Schematic representation of MgII and FeII emission regions. The mean distance of the MgII and FeII region from the BH is $\sim$0.23 pc, estimated from the rest frame time delays, 276 days and 270 days respectively for MgII and FeII. The FeII emission also exhibits a smaller time-delay in the rest frame, 180 days, which indicates the larger extent of the FeII region with respect to the MgII emission region in the direction away from the SMBH. Hence, due to a non-negligible inclination of the observer, eventually the reprocessed emission from the FeII region tends to come to the observer sooner. The secondary time delays at $\sim 570$ and $\sim 680$ days can potentially be attributed to the ``mirror'' effect, i.e. the MgII and FeII emission, respectively, coming to the observer from the other side of the accretion disc, which is supported by the temporal difference of $(570-270)=300$ days that corresponds to $\sim 0.25\,{\rm pc}$, i.e. one additional disc crossing of photons. However, the FeII emission coming from the more distant part across the disc is partially shielded by the MgII line-emitting region closer in with respect to the observer. Note- color do not represent the density of the cloud.}
    \label{fig:cartoon}
\end{figure*}

\section{Discussion}
\label{sec_discussion}

We determined the time delay of the MgII emission and FeII emission with respect to the continuum in the quasar CTS C30.10 (z=0.90052). While MgII time delay was already determined for several sources (see \citealt{khadka2021} for a current list of sources), FeII time delays in the UV were not generally measured. So far UV FeII  time delay was reported only for NGC 5548, and \citet{Maoz1993} found a similar time lag for the UV FeII lines as for the Ly$\alpha$.

Using several methods, we determined the MgII time delay at  $523.5^{+23.6}_{-37.1}$ days in the observed frame. The time delay for Fe II was not unique, with two values favored: $513.0^{+26.2}_{-48.1}$ and $342.5^{+50.6}_{-57.0}$ days. This likely suggests a more complex reprocessing region. In some cases it is
clear that Fe II has more than one component and is well fitted with two components (\citealt{dong2010}, \citealt{hryniewicz2022}).

The time delays for FeII seem on average by some 10\% shorter than for MgII, for a given method, and indication of much shorter time delay appears.  This is rather unexpected at the basis of the kinematic line width. The FeII template used in the spectral fitting was broadened to FWHM = 2115 km$s^{-1}$, This broadening was adjusted at the basis of $\chi^2$ optimization by \citet{Modzelewska2014}, and tested also with the current data. The MgII line was fitted as a two-component line, and the mean value of the FWHM of component 1 is $2756 \pm 122$ km $s^{-1}$, for the component 2 is $3558 \pm 102 $km $s^{-1}$, and if the line is actually treated as a single component line of more complex shape, the total FWHM is $4868 \pm 114$ km $s^{-1}$. Therefore, the line width implies the location of FeII further out in comparison to MgII, consistent with the ionization potential being lower for FeII. 

Therefore we also used a wavelength-resolved approach to a signal containing both MgII and FeII, with the aim to shed more light onto the relative geometry of the two regions. The data quality is not excellent but we see similar overall trend as for the MgII and FeII lightcurves. In Table~\ref{tab_totalMgII_FeII} and Table~\ref{tab:curves1_7} we give the results for peak as well as median values for a given method. In the case of ICCF, the discussion by \citet{koratkar1991} shows that the centroid-based values correspond to the luminosity-weighted radius while peak values are more affected by the gas at small radii. However, our ICCF-based results do not show significant differences there, being consistent rather with a relatively compact reprocessing region.

To test independently the extension of the emitting regions we calculated the Auto-correlation function (ACF) of total MgII and FeII along with continuum. The result is show in Figure \ref{fig:ACF}. The ACF of continuum decays at timescales 250 days. The secondary peak reappears at timescale of 750 days. In their central parts, ACF of FeII is broader than the MgII suggesting more extended emission region for FeII. Both functions, however, show a form of a plateau at timescales 200 - 500 days, which is not a typical feature of AGN lightcurves. All these unexpected effects are likely connected to the apparent similarity of two peaks in the continuum lightcurve, separated by $\sim 1500$ days. In still longer data sequence this apparent similarity would likely disappear. 

Both MgII and FeII emission regions are however extended, and the apparent discrepancy between the narrower FeII lines but shorter effective FeII delay can be solved as illustrated in Figure \ref{fig:cartoon}. If the observer is inclined with respect to the symmetry axis, there is a clear asymmetry in the visibility of the BLR part closer to the observer and the part located at the other side of the black hole. The region emitting mostly MgII can be still transparent for the continuum, so FeII is produced, but FeII emitting region can partially suppress the MgII emission from the near side. The opposite can happen for the farther side - now MgII emitting region is much better exposed while FeII is partially shielded. Since the measured time delay is the weighed average over the entire region, the net FeII time delay can still be shorter since the closest part dominates more.

\subsection{BLR kinematics}

The wavelength-resolved or velocity-resolved time lags allow us to explore the BLR geometry and kinematics. The distribution of estimated time lags with corresponding wavelength can have different shape namely, symmetric shape suggesting the Keplerian or disk-like rotation of BLR. 
Results obtained by other authors for H$\beta$ line which also belongs to low ionization lines, like MgII and FeII bring different results. \citet{Grier_2017} from their study of four sources did not claim any outflows/inflows although requested the presence of the elliptical orbits. \\
Velocity-resolved reverberation mapping of 3C120, Ark120, Mrk 6, and SBS 1518+593 was done by \citet{Du2018}. Their results show that the first three AGN has a complex features different from the simple signatures expected for pure outflows, inflow, or Keplerian disk. Moreover, SBS 1518+593 show least asymmetric velocity-resolved time lags characteristic of a Keplerian disk. They also observed a significant change in the velocity-resolved time lags of 3C 120 compared to its previous study, suggesting an evolution of BLR structure. 
\citet{Hu_2020} have studied the quasar PG 0026+129 and their results shows an evidence of two distinct broad line region (BLR). Their velocity-resolved analysis supports two regions but does not imply any simple inflow/outflow pattern.
\citet{Lu_2016} provides the detail study of reverberation mapping of the broad line region in NGC 5548. Their velocity-binned delay map for the broad H$\beta$ line shows a symmetric response pattern around the line center. They suggest it could be a plausible kinematic signature of virialized motion of the BLR.
Another study of NGC 5548 by \citet{Pei_2017}  found a complex velocity-lag structure with shorter lags in the line wings of H$\beta$. They concluded again that the broad line region is dominated by Keplerian motion. The same conclusion, for the same source, was reached by \citet{Xiao_2018}. 
\\
Recently, \citet{U2021} have reported the velocity-resolved H$_{\beta}$ time lags for several nearby bright Seyfert galaxies where they have observed all possible scenario including the Keplerian motion of BLR and the radially in-falling and out-flowing materials.
A study of high ionization line velocity structure (CIV) in NGC 5548 has been done by \citet{Rosa_2015}, with the six month long observation taken from Cosmic Origins Spectrograph on the Hubble Space Telescope. They observed a significant correlated variability in the continuum and the broad emission lines. Their velocity-resolved time lag study shows coherent structure in lag versus line of sight velocity for the emission lines, but they do not see clear outflow signatures. This could be related to relatively small Eddington ratio in this source, and no clear shift in CIV, frequently seen in quasars. \\
In our data we do not see any monotonic increase or a decrease of the time delay with the wavelength which are the signatures of the inflow or the outflow. Thus the dynamics in CTS C30.10 seems to be consistent with the predominantly Keplerian motion. Time delays measured in a wavelength-dependent way for a combination of MgII and FeII as well as separate time delay measurement for total MgII and FeII emission, combined with kinematic line width, imply a stratification in the BLR, but also a clear asymmetry in the visibility of the BLR part closer to the observer and the part which is more distant, as visualized in Figure~\ref{fig:cartoon}.

\subsection{Updated MgII radius-luminosity relation}

\begin{figure}[h!]
    \centering
    \includegraphics[width=\columnwidth]{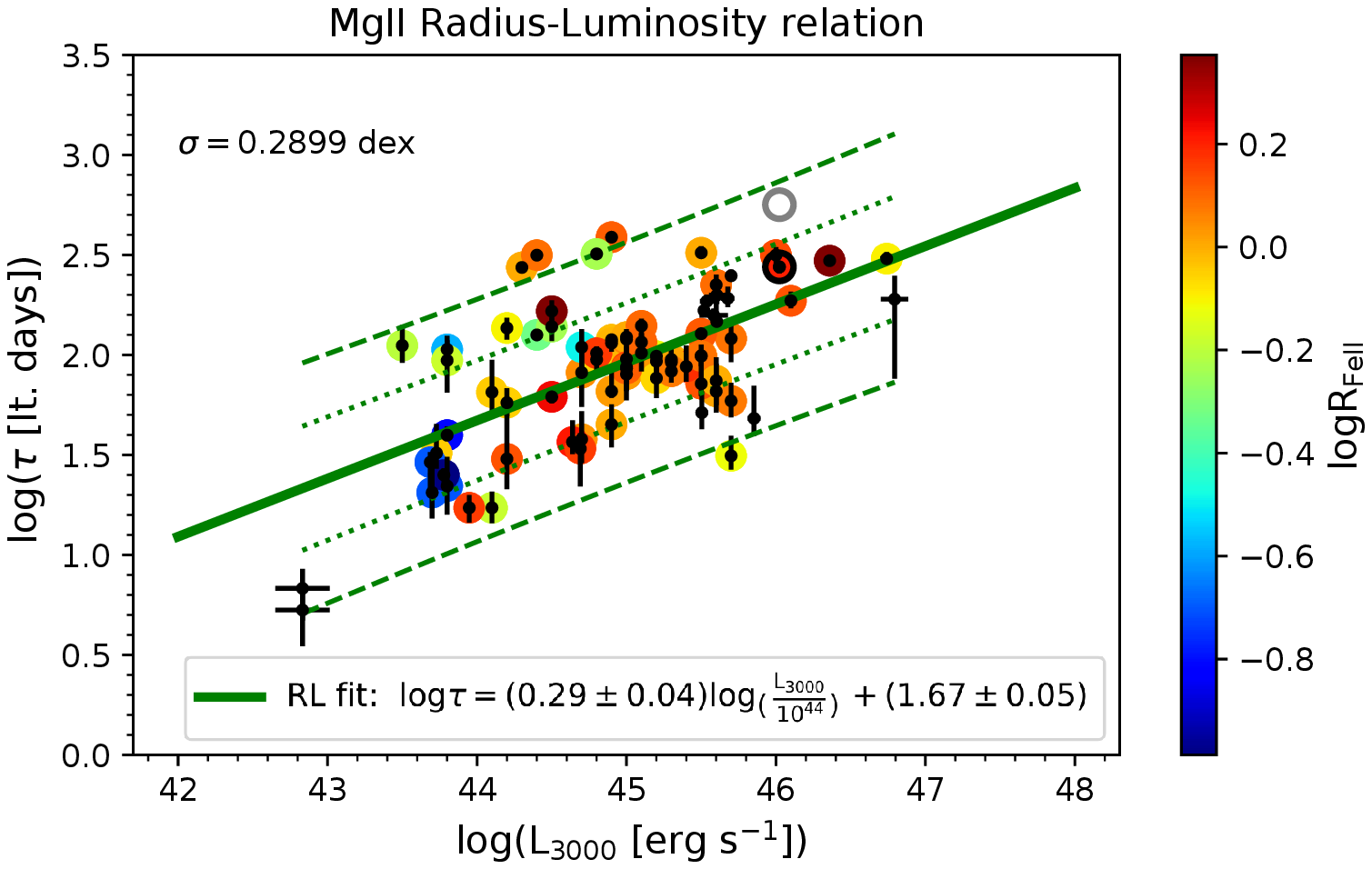}
    \caption{The radius-luminosity relation for the currently available 78 MgII sources. The best-fit relation determined by the classical least-square fitting is indicated by the solid green line. The updated MgII emission-line time delay of $275.5^{+12.4}_{-19.5}$ days is depicted by a black circle, while the old measurement is shown as a gray circle for comparison. The overall scatter is $\sim 0.29$ dex. For 66 sources, we have available the measurements of the relative FeII strength ($R_{\rm FeII}$ parameter), which are colour-coded according to the axis on the right. The dotted and dashed green lines show 1 and 2$\sigma$ prediction intervals, respectively, for the sample of 78 MgII RM sources.}
    \label{fig_RL_relation}
\end{figure}

\begin{figure*}
    \centering
    \includegraphics[width=0.49\textwidth]{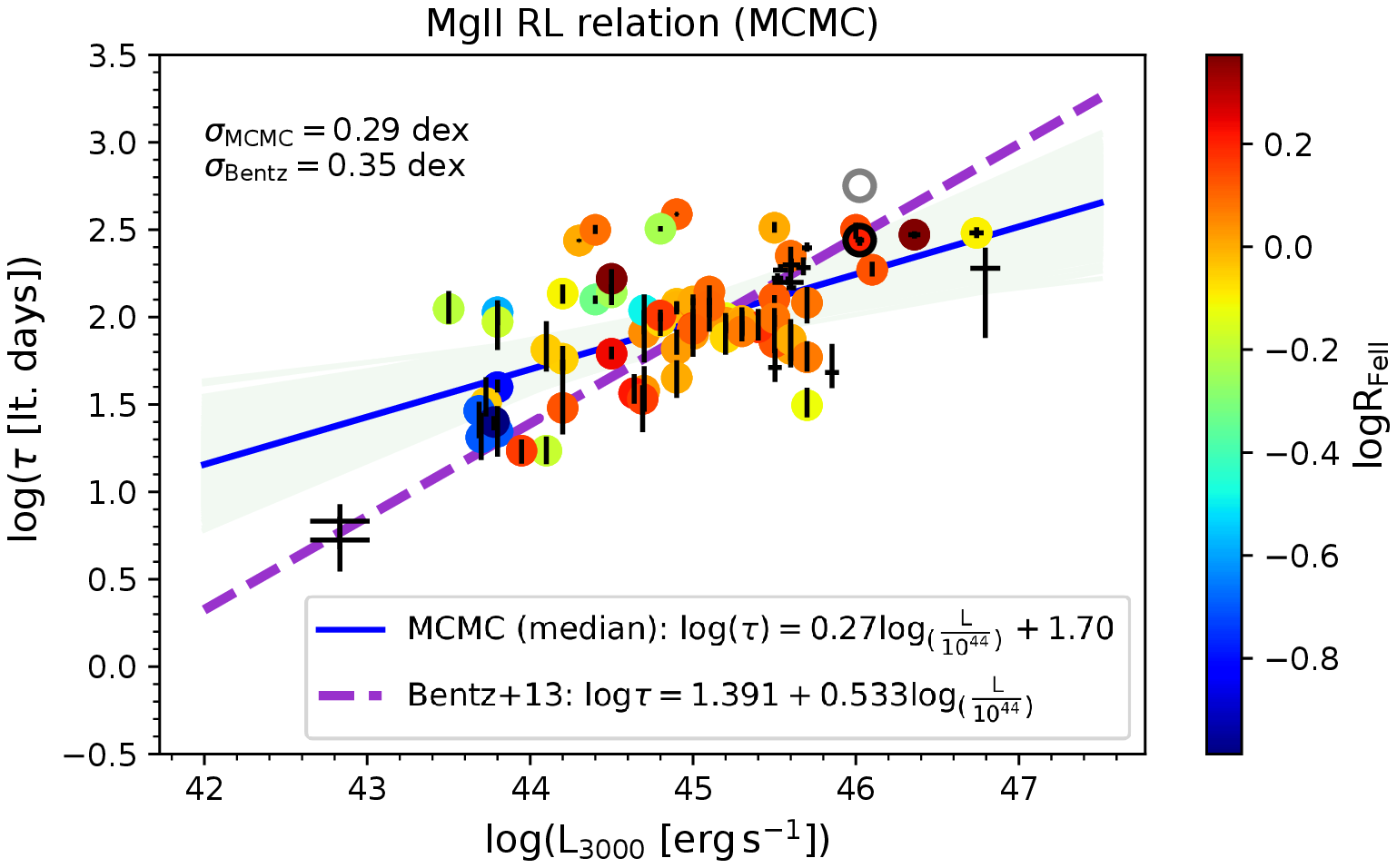}
    \includegraphics[width=0.49\textwidth]{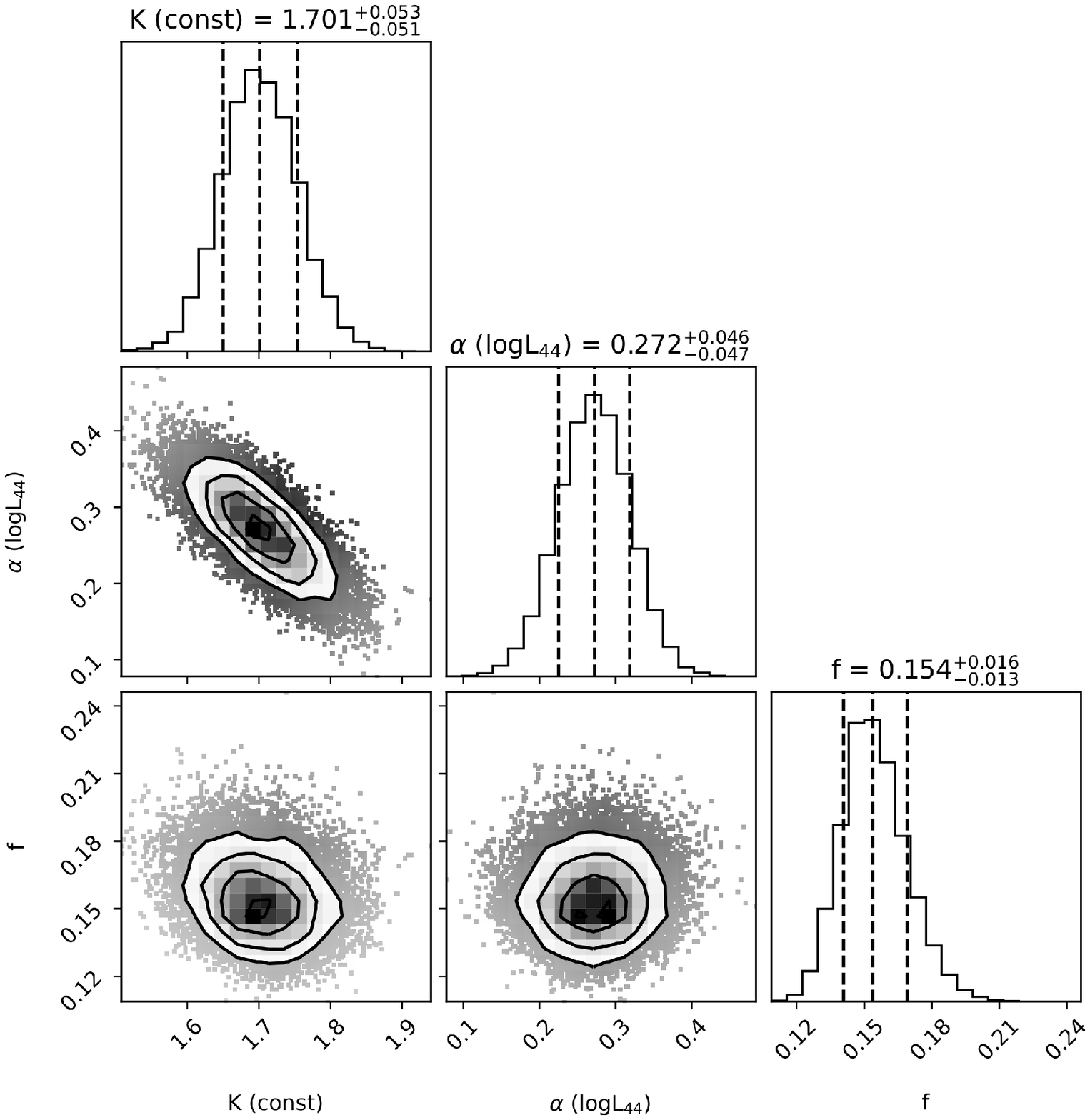}
    \caption{The MgII maximum-likelihood radius-luminosity relation determined using the Markov-Chain Monte Carlo algorithm (left panel). The coefficients as well as the RMS scatter are consistent within uncertainties with the classical fitting algorithm (see Figure~\ref{fig_RL_relation}). CTS C30.10 is depicted using the black circle, while the older time-delay measurement, approximately twice as long, is represented by a gray circle. The dashed violet line shows the Bentz RL relation with the slope of $\sim 0.5$, which was adjusted for the continuum luminosity at $3000\,\AA$. In the right panel, we show the corner plot representation of the distribution histograms for the two parameters $\alpha$ (slope) and $K$ constant in the linear fit of $\log{\tau}=\alpha \log{L_{44}}+K$ to the MgII data. The likelihood function included the underestimation factor $f$ whose distribution is also shown.}
    \label{fig_RL_relation_mcmc}
\end{figure*}

Following our previous constructions of the MgII-based radius-luminosity (RL) relation \citep{Czerny2019,Zajacek2020,2020ApJ...903...86M,zajacek2021,khadka2021}, we update this relation for 78 available reverberation-mapped sources, including the updated rest-frame time delay of the total MgII emission for CTS C30.10, $\tau_{\rm MgII}=275.5^{+12.4}_{-19.5}$ days, see Figure~\ref{fig_RL_relation} and \ref{fig_RL_relation_mcmc}. The RL relation is generally well defined with a significant positive correlation between the rest-frame MgII time delay and the monochromatic luminosity at 3000\AA. The Spearman correlation coefficient is $s=0.49$ ($p=4.96 \times 10^{-6}$) and the Pearson correlation coefficient is $r=0.63$ ($p=8.39\times 10^{-10}$), which motivates the search for a power-law relation in the form $\tau=KL^{\alpha}$.

We fit the linear function $\log{\tau}=\alpha \log{(L_{3000}/10^{44}\,{\rm erg\,s^{-1}})}+K$ to the 78 MgII data using the classical least-square fitting procedure as well as the Monte-Carlo Markov Chain algorithm. From the least square fitting, we obtain the best-fit radius-luminosity relation,
\begin{equation}
    \log{\left(\frac{\tau}{1 \text{lt. day}} \right)}=(0.29 \pm 0.04)\log{\left(\frac{L_{3000}}{10^{44}\,{\rm erg\,s^{-1}}} \right)}+(1.67\pm 0.05)\,,
    \label{eq_RL_least_square}
\end{equation}
with the scatter of $\sigma=0.29$ dex. The best-fit relation is depicted in Figure~\ref{fig_RL_relation} along with 78 RM sources, 66 of which are colour-coded according to the relative FeII strength, $R_{\rm FeII}$, which serves as a suitable observational proxy for the accretion-rate intensity \citep{2020ApJ...903...86M}.

Using the MCMC algorithm, including the uncertainty underestimation factor $f$, we obtain the maximum-likelihood radius-luminosity relation,
\begin{equation}
    \log{\left(\frac{\tau}{1 \text{lt. day}} \right)}=(0.27^{+0.05}_{-0.05})\log{\left(\frac{L_{3000}}{10^{44}\,{\rm erg\,s^{-1}}} \right)}+(1.70^{+0.05}_{-0.05})\,,
    \label{eq_RL_MCMC}
\end{equation}
with the scatter of $\sigma\simeq 0.29$ dex. Both the RMS scatter as well as the inferred radius-luminosity relation are consistent within the uncertainties with the values determined from the least-square fitting.
The maximum-likelihood relation is shown along with the MgII data in Figure~\ref{fig_RL_relation_mcmc} (left panel) alongside the corner plot (right panel) with the slope and the intercept distributions.
The new shorter MgII time delay (depicted by a black circle in Figure~\ref{fig_RL_relation} and \ref{fig_RL_relation_mcmc}) now lies within 2$\sigma$ prediction interval of the whole sample, see Figure~\ref{fig_RL_relation}, while previously it was within the 1$\sigma$ interval. Also for the MCMC fitting, the new MgII time-delay lies within 2$\sigma$ of the median RL relation. 

In addition, we also consider for comparison the former RL relation based on the reverberation-mapped H$\beta$ sample \citep{Bentz2013}. The Bentz relation has a slope of $\sim 0.5$, i.e. consistent with the simple photoionization theory. We transform the Bentz relation inferred for the monochromatic luminosity at $5100\,\AA$ to the RL relation for $3000\,\AA$ and obtain $\log{\tau}=1.391+0.533\log{(L_{3000}/{10^{44}\,{\rm erg\,s^{-1}})})}$, i.e. the same slope but a slightly smaller intercept \citep{Zajacek2020}. Interestingly, the smaller time delay of CTS C30.10 is now consistent with this relation, see Figure~\ref{fig_RL_relation_mcmc} (left panel). However, the RMS scatter of the whole MgII sample along the Bentz relation is larger ($\sigma\sim 0.35\,{\rm dex}$) in comparison with the maximum-likelihood RL relation ($\sigma\sim 0.29\,{\rm dex}$). 

In \citet{2019ApJ...886...42D}, the authors investigated the extended radius-luminosity relationship using the relative FeII strength with respect to the H$\beta$ line, $R_{\rm FeII}=\text{EW(FeII)}/\text{EW(H$\beta$)}$. The idea was to decrease the scatter of H$\beta$ sources along the RL relation, which appears to be driven by the accretion-rate intensity. Since $R_{\rm FeII}$ is correlated with the accretion rate, it should provide the correction. Indeed, a certain improvement was reported with the final scatter of $0.196$ dex in comparison with the original H$\beta$ RL relation with $\sigma \sim 0.28$ dex \citep{2018ApJ...856....6D}.

In our case, the relative strength is defined analogously as the ratio of the equivalent width of FeII to the equivalent width of MgII, $R_{\rm FeII}=\text{EW(FeII)}/\text{EW(MgII)}$. Hence, we investigate analogously to \citet{2019ApJ...886...42D} the extended RL relation in the form $\log{\tau}=\alpha \log{(L_{3000}/10^{44}\,{\rm erg\,s^{-1}})}+\beta\log{R_{\rm FeII}}+K$. The correlation between the rest-frame time delay $\log{\tau}$ and the term $\log{L_{\rm 44}}+\beta/\alpha\log{R_{\rm FeII}}$ \footnote{$L_{\rm 44}\equiv L_{3000}/(10^{44}\,{\rm erg\,s^{-1}})$.} is weaker than for the simple RL relation, but still present, with the Spearman correlation coefficient of $s=0.39$ ($p=0.0013$) and the Pearson correlation coefficient of $r=0.52$ ($p=6.16 \times 10^{-6}$).  

\begin{figure*}
    \centering
    \includegraphics[width=0.49\textwidth]{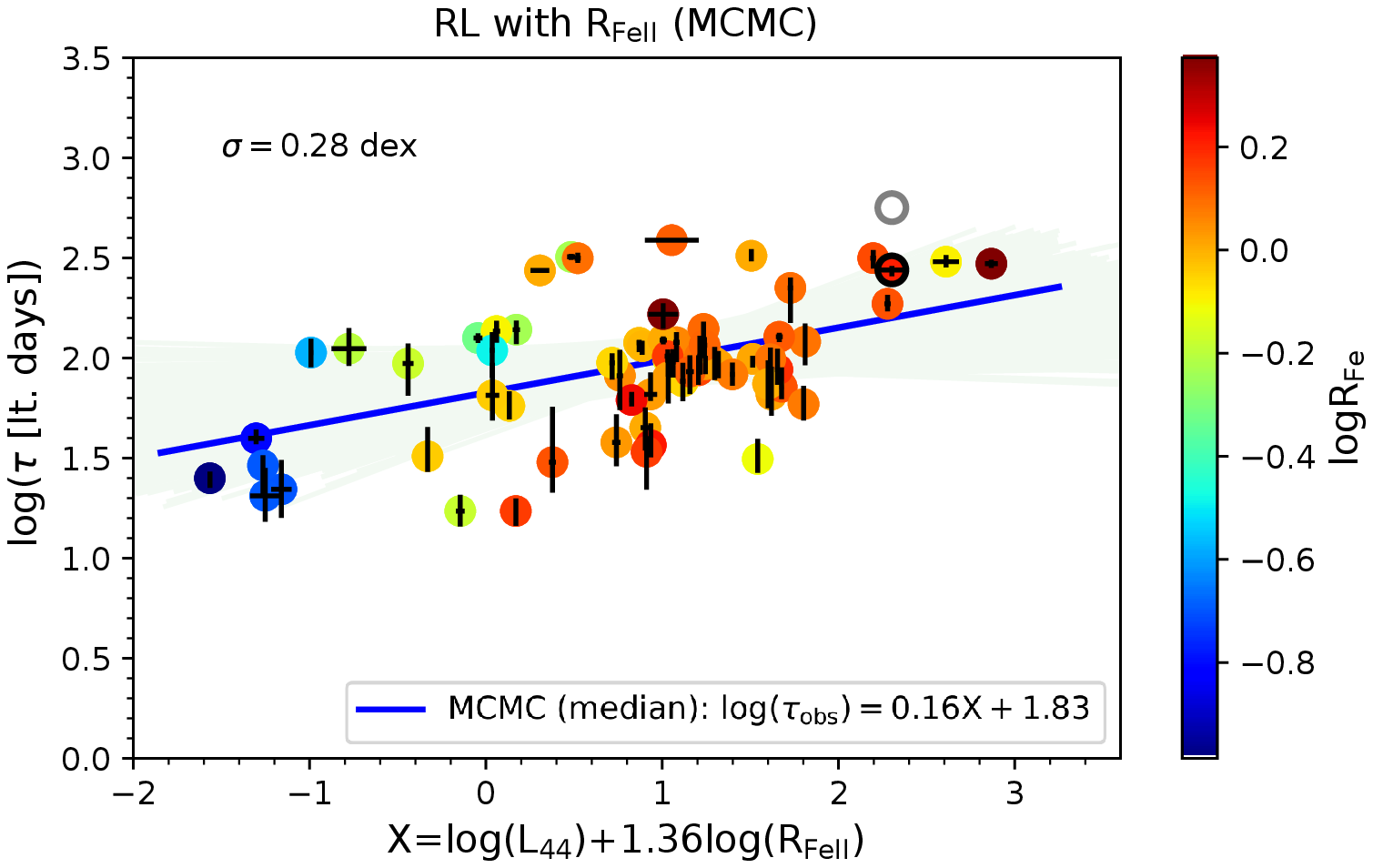}
    \includegraphics[width=0.49\textwidth]{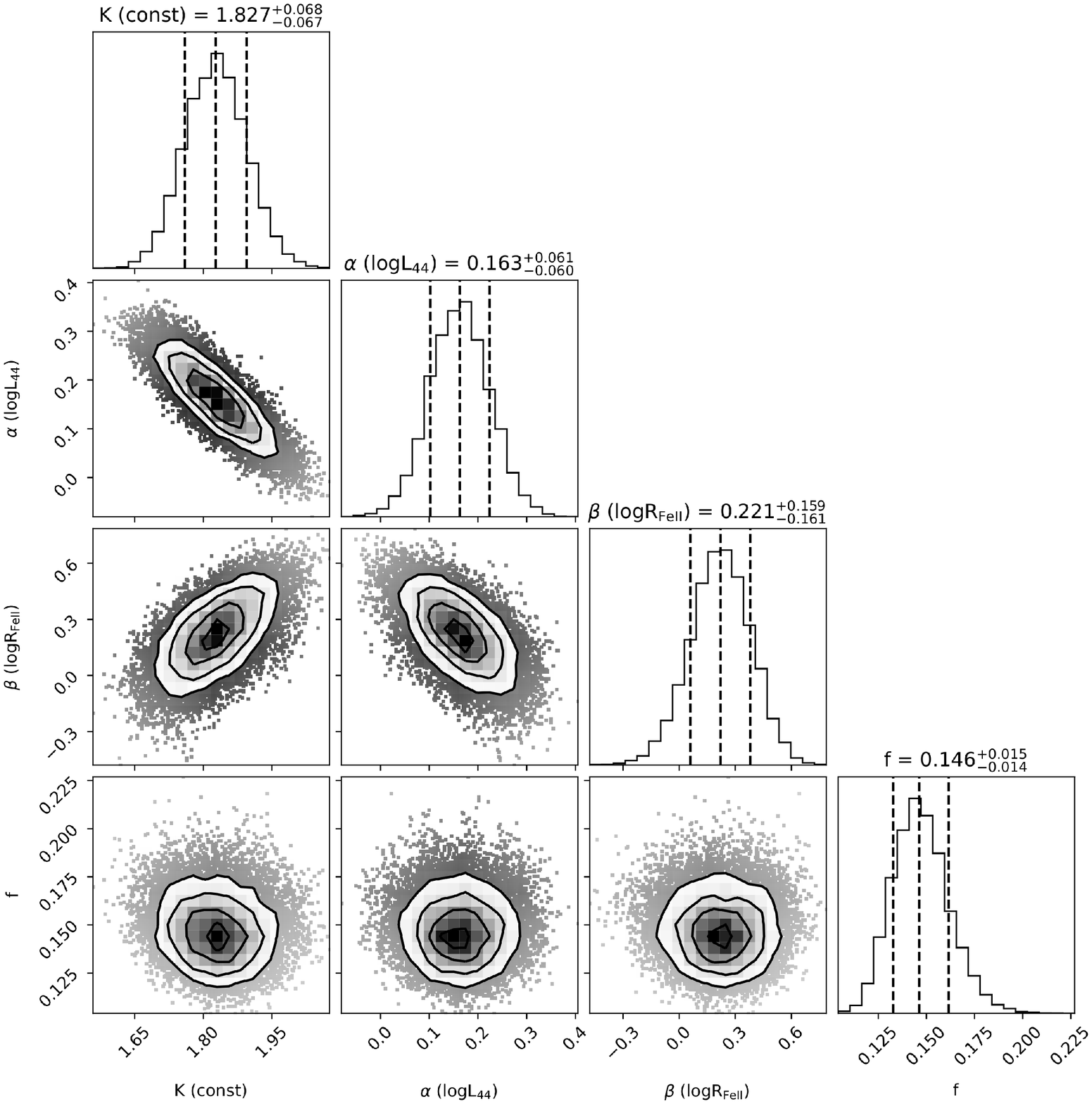}
    \caption{The extended radius-luminosity relation including the relative FeII strength, $R_{\rm FeII}$. In the left panel, the median relation (blue line) is inferred by maximizing the likelihood function, green lines show 1000 random selections from the parameter distribution. 66 MgII time-delay measurements are colour-coded to depict $\log{R_{\rm FeII}}$ according to the colour axis on the right. The RMS scatter is $0.28$ dex. In the right panel, the corner plot shows the parameter distributions inferred from the MCMC method. The parameter uncertainties correspond to $16\%$ and $84\%$ percentiles. }
    \label{fig_RL_relation_RFe_mcmc}
\end{figure*}

In Figure~\ref{fig_RL_relation_RFe_mcmc} (left panel), we show the extended RL relation for 66 sources with available $R_{\rm FeII}$ measurements, with an updated time-delay of CTS C30.10, $\tau_{\rm MgII}=275.5^{+12.4}_{-19.5}$ days, including the median RL relation inferred using the MCMC method as well as the distribution of 1000 randomly selected relations. In the right panel of Figure~\ref{fig_RL_relation_RFe_mcmc}, we display the corner plot representation of parameter distributions. Maximizing the likelihood function leads to the following parameter inferences: $\alpha=0.163^{+0.061}_{-0.060}$, $\beta=0.221^{+0.159}_{-0.161}$, and $K=1.827^{+0.068}_{-0.067}$. The underestimation factor is $f=0.146^{+0.015}_{-0.014}$. The RMS scatter is $0.28$ dex, which is a decrease only by $3.7\%$ with respect to a simple RL relation with the scatter of $0.29$ dex (displayed in Figure~\ref{fig_RL_relation_mcmc}). Hence, we cannot confirm a significant decrease of the scatter for MgII sources for the extended RL relation. However, splitting the source sample into high- and low-accretors can lead to a significant scatter decrease, at least for the current sample of RM MgII quasars, which was analyzed by \citet{2020ApJ...903...86M}.

\subsection{UV FeII radius-luminosity relation}

Previously only one measurement of the UV FeII pseudocontinuum was performed for the Seyfert galaxy NGC5548 during the campaign in 1988-1989 with IUE satellite \citep{Maoz1993}. The FeII time delay centroid was inferred to be $\tau_{\rm NGC5548}^{\rm FeII}=10\pm 1$ days. The corresponding monochromatic luminosity at 3000\,\AA\, for this source is  $\log{[L_{3000}\,(\text{erg\,s$^{-1}$})]}=43.696 \pm 0.051$ according to the NED database\footnote{https://ned.ipac.caltech.edu/}, which was inferred considering the flux densities determined close to $3000\,\AA$. 

\begin{figure}
    \centering
    \includegraphics[width=\columnwidth]{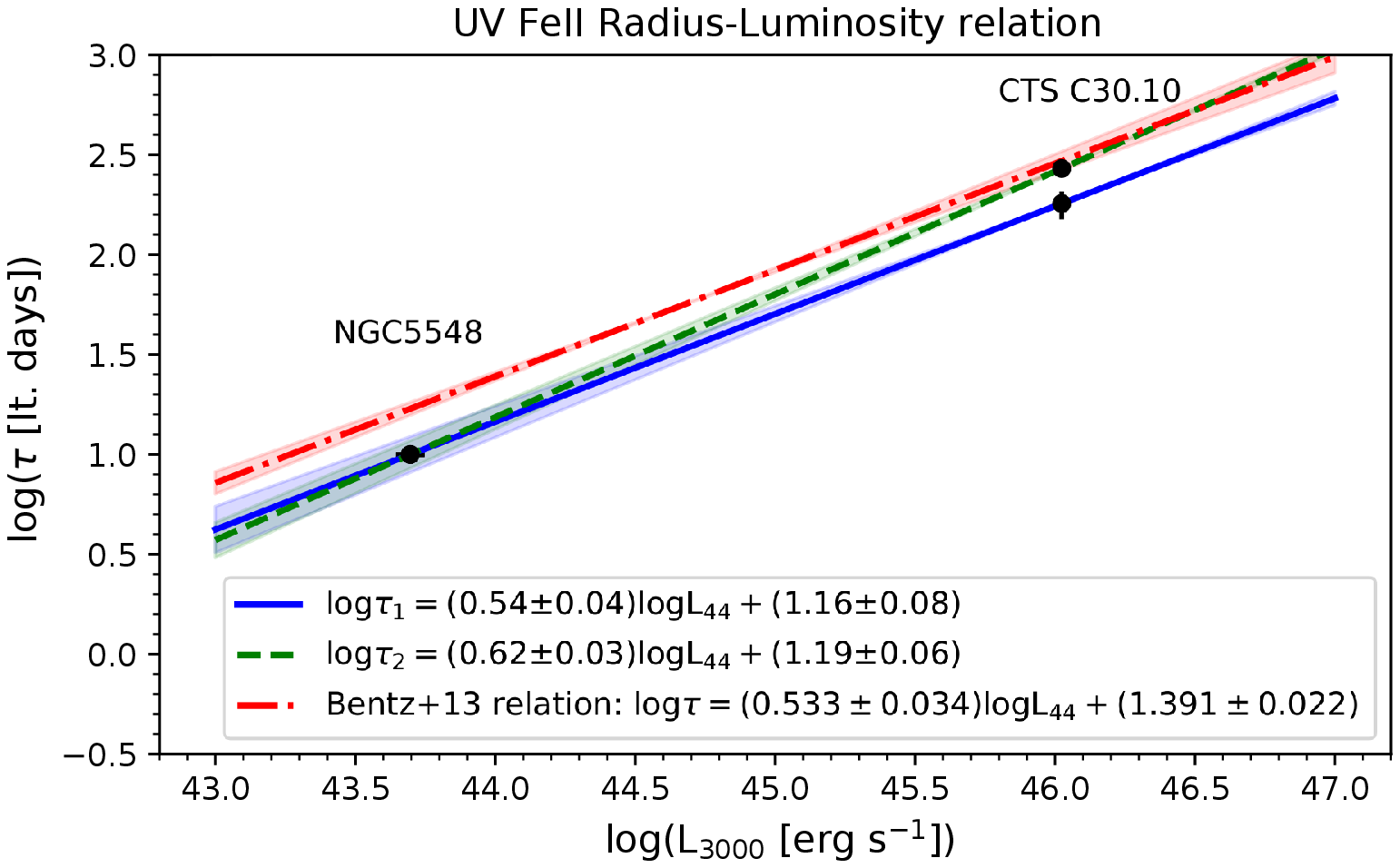}
    \caption{The preliminary radius-luminosity relation for the UV FeII pseudocontinuum based on the two time-lag measurements for NGC5548 \citep{Maoz1993} and CTS C30.10 (this work). We derive two relations based on the two FeII time delays for CTS C30.10: $180.3$ days and $270.0$ days. The larger time delay of $270.0$ days is more consistent with the RL relation derived by \citet{Bentz2013} when renormalized to the 3000\,\AA\, monochromatic luminosity. The shaded regions depict $1\sigma$ confidence intervals of individual RL relations.}
    \label{fig_feII_RL}
\end{figure}

Here we report two potential time-delay peaks for the UV FeII pseudocontinuum for a more luminous source CTS C30.10, $\tau_{\rm CTS1}^{\rm FeII}=180.3^{+26.6}_{-30.0}$ days and $\tau_{\rm CTS2}^{\rm FeII}=270.0^{+12.4}_{-19.5}$ days. The monochromatic luminosity for the source is $\log{[L_{3000}\,(\text{erg\,s$^{-1}$})]}=46.023 \pm 0.026$. Since the UV FeII radius-luminosity relation was not investigated before, the two measurements across three orders of magnitude in luminosity now allow us to make a preliminary discussion of its existence for the first time.

In Figure~\ref{fig_feII_RL}, the preliminary UV FeII radius-luminosity relation is outlined, which confirms the basic trend and the power-law dependency with the slope close to $0.5$. Using the smaller FeII time delay of 180.3 days for CTS C30.10 yields the relation $\log{\tau}=(0.54\pm 0.04)\log{L_{44}}+(1.16 \pm 0.08)$, while the larger time delay of $270.0$ days leads to $\log{\tau}=(0.62 \pm 0.03)\log{L_{44}}+(1.19 \pm 0.06)$. The uncertainties were calculated by the propagation of errors of the time delays and the luminosities of the two sources. The longer time delay for FeII is more consistent with the radius-luminosity relation of \citet{Bentz2013} when renormalized for the monochromatic luminosity at 3000\AA. This is in agreement with the picture sketched in Figure~\ref{fig:cartoon}, where MgII and FeII line-emitting regions share approximately the common mean distance from the SMBH corresponding to $\sim 270-275.5$ days. The FeII emitting region is, however, more extended towards larger distances from the SMBH, with one side closer to the observer, which produces effectively the second, shorter time-delay peak. 

\section{Conclusions}
\label{sec_conclusions}

We summarize the main results of nine years of the monitoring of the luminous quasar CTS C30.10 (2012-2021) as follows:
\begin{itemize}
    \item MgII line-emission exhibits a rest-frame time-delay of $275.5^{+12.4}_{-19.5}$ days, which is about a factor of two less than previously reported value for this source. This shows that the duration of the monitoring is essential to accurately determine the emission-line time delay, especially if more time-delay peaks are present,
    \item the MgII time-delay is consistent within $2\sigma$ with the best-fit RL relation for all current RM MgII quasars. It also lies on the previously determined H$\beta$ RL relation with the slope close to $0.5$,
    \item the rest-frame time-delay for the FeII emission has two components: $270.0^{+13.8}_{-25.3}$ days and $180.3^{+26.7}_{-30.0}$ days. Since the FeII line width is smaller than for the MgII line, it is expected be located further from the SMBH. On the other hand, the mean distance of the emission regions is comparable, as is indicated by the common time-delay component within uncertainties. The shorter time-delay component indicates that the observer predominantly sees the FeII emission region oriented towards them, while the more distant region is mostly shielded by the closer MgII region,
    \item combining our UV FeII time-delay measurement with the older one in NGC5548, we find that these measurements point towards the existence of the UV FeII radius-luminosity relation, whose slope is consistent with $0.5$ within uncertainties. The longer FeII time-delay is consistent with the H$\beta$ RL relation, which indicates that the time-delay of $\sim 270$ days expresses the mean distance of the FeII region, while the shorter time delay of $\sim 180$ days is associated with the extension of FeII region closer to the observer and away from the SMBH, 
    \item the wavelength-resolved reverberation mapping of the MgII$+$FeII complex between $2700$ and $2900\,\AA$ shows that this region is stratified with the core of the MgII emission line having a larger time delay with respect to the wings dominated by FeII, which reflects the geometrical orientation of this complex with respect to the observer.   
\end{itemize}

\begin{acknowledgements}
This paper uses observations made at the South African Astronomical Observatory (SAAO).
The project is based on observations made with the
SALT under programs 2012-2-POL-003, 2013-1-POL-RSA-
002, 2013-2-POL-RSA-001, 2014-1-POL-RSA-001, 2014-2-
SCI-004, 2015-1-SCI-006, 2015-2-SCI-017, 2016-1-SCI-011,
2016-2-SCI-024, 2017-1-SCI-009, 2017-2-SCI-033, 2018-1-
MLT-004 (PI: B. Czerny). 
      The project was partially supported by the Polish Funding Agency National Science Centre, project 2017/26/A/ST9/00756 (MAESTRO 9), and MNiSW grant DIR/WK/2018/12.
      Some of the observations reported in this paper were obtained with the Southern African Large Telescope (SALT). Polish participation in SALT is funded by grant No. MNiSW DIR/WK/2016/07. The authors also acknowledge the Czech-Polish mobility program (M\v{S}MT 8J20PL037 and
NAWA PPN/BCZ/2019/1/00069). 
      MZ acknowledges the financial support of the GA\v{C}R EXPRO grant No. 21-13491X ``Exploring the Hot Universe and Understanding Cosmic Feedback". SP acknowledges financial support from the Conselho Nacional de Desenvolvimento Científico e Tecnológico (CNPq) Fellowship (164753/2020-6).
\end{acknowledgements}

%
%
\bibliographystyle{aa}
\bibliography{output.bib}
\begin{appendix}

\section{Methods}

\subsection{Time lag Measurements}
\label{subsec_timelag_methods}
We have applied various time-delay measurement methods to assess a time delay between the continuum and the various wavebands of the emission light curve. This way we minimize the bias of the individual methods. 

\subsubsection{ICCF}
\texttt{ICCF} stands for the interpolated cross-correlation function that is frequently used to determine the time lag between continuum and the emission lines in quasars. A detailed description of the ICCF is provided in \citet{Gaskell1987} and \citet{Peterson1998, Peterson2004}. \texttt{ICCF} first interpolates the light curves and then estimates the time lags among the curves. We have used the python version of the \texttt{ICCF}, i.e. \texttt{pyCCF} developed by \citet{Sun2018}. The result of \texttt{pyCCF} shows a broader peak in time lags. To estimate the best time lag, we first estimate the centroid and the peak of the distribution and the time lag corresponding to the "median" of the distribution is 
considered to be the best time lag. To obtain the uncertainties of the time lags, we have followed the flux randomization (FR) and the random subset selection (RSS) technique discussed in \citet{Maoz1990}, \& \citet{Peterson1998, Peterson2004}. FR and RSS methods randomizes the observed flux with respect to their uncertainties and re-sample the light curves. The procedure has been followed for 10,000 realizations and eventually the CCF was estimated. The centroid and the peak of each CCF run forms a cross-correlation centroid distribution (CCCD) and a cross-correlation peak distribution (CCPD). Further, the value of the time lag and its uncertainty was estimated from the CCCD/CCPD from its 84.13$\%$ quantiles.

\subsubsection{Javelin}
\texttt{JAVELIN} stands for "Just Another Vehicle for Estimating Lags In Nuclei". The method is very commonly used for the reverberation mapping in quasars. The detailed description of the method can be found in \citet{Zu2011}\footnote{ \href{https://github.com/nye17/javelin}{Javelin}}. 
It models the continuum light curve using the damped random walk (\citealt{Kelly2009}, \citealt{2010ApJ...708..927K}, \citealt{Zu2013}) procedure to estimate the time lags between continuum and emission lines. Before estimating the time lags, \texttt{JAVELIN} also models the emission-line light curve as a smoothed, scaled, and a lagged version of the continuum line curve. To determine the best time lag and its uncertainty, we applied the bootstrap method and the procedure is followed for 1000 realizations.

\subsubsection{$\chi^2$}
The $\chi^2$ method is another robust technique to detect potential time lags in quasars. Previously, it has mainly been used for detecting time lags caused by lensing in quasars. In fact, a comprehensive study by \citet{Czerny2013} suggests that the $\chi^2$ method works more reliably than the \texttt{ICCF} method in case of the red-noise dominated AGN variability. The $\chi^2$ procedure is similar to the \texttt{ICCF}, where one of the light curves is shifted with respect to the other one, and the $\chi^2$ minimization technique is used to find the similarity between the shifted curve and the original curve. The shifted time corresponding to the minimum $\chi^2$ value is potentially the time delay between the two curves (continuum and emission line). To estimate the error of time lags, the bootstrap procedure has been followed as described for \texttt{JAVELIN}.

\subsubsection{Measures of data regularity/randomness: von Neumann and Bartels estimators}

To estimate the best time delay between the continuum and the MgII or the FeII line emission, we applied the measures of data regularity/randomness \citep{2017ApJ...844..146C} that have been previously applied in cryptography or data compression. The estimators of data regularity make use of the unified light curve that is constructed from the continuum light curve $F_1$ and the time-shifted line-emission light curve $F_2^{\tau}$: $F(t,\tau)=\{t_i,f_i\}_{i=1}^{N}=F_1 \cup F_2^{\tau}$, where $N=N_1+N_2$ is the sum of light curve data points. In particular, the optimized von Neumann's estimator for a time delay $\tau$, $E(\tau)$, is defined as the mean of the squared successive differences of $F(t,\tau)$,
\begin{equation}
    E(\tau)=\frac{1}{N-1}\sum_{i=1}^{N-1}[F(t_i)-F(t_{i+1})]^2\,.
    \label{eq_von_neumann}
\end{equation}
The minimum of $E(\tau')$ corresponds to the time delay $\tau'$ for which $F(t,\tau)$ is the most `regular', i.e. its power spectrum is dominated by long-term changes while short-term effects are effectively suppressed. In other words, for the minimum of $E(\tau)$, the combined light curve resembles a red-noise process rather than a white-noise variability, and the time delay $\tau'$ may be considered as a good estimate of the true time delay, $\tau'\sim \tau_0$. The Bartels estimator is similar to the optimized von Neumann's scheme, but it makes used of the ranked version of the combined light curve $F_{\rm R}(t,\tau)$. The advantage of the measures of data regularity is that they do not introduce a bias to the data via the polynomial interpolation (as the ICCF and $\chi^2$ methods do), the binning in the correlation space (as is performed by the DCF and the zDCF), or the modelling of the continuum variability (e.g. using the damped random-walk process that is used by the JAVELIN).

\subsubsection{DCF}

The correlation studies using the discrete correlation function (DCF) formulated by 
\citet{Edelson1988} is also used to estimate the time lags.
If we have two discrete data sets a$_i$ and b$_j$ with the standard deviations $\sigma_a$ and $\sigma_b$, the discrete correlation coefficient for all the measured pairs (a$_i$-b$_j$) is defined as,
\begin{equation}
 UDCF_{ij} = \frac{(a_i-\bar{a})(b_j-\bar{b})}{\sqrt{(\sigma_a^2-e_a^2)(\sigma_b^2-e_b^2)}}\,,
\end{equation}
where each pair is associated with a pairwise lag $\Delta t_{ij}$ = t$_j$ - t$_i$. The measurement errors associated with data sets a$_i$ and b$_j$ are denoted as e$_a$ and e$_b$, respectively. Averaging the UDCF$_{ij}$ over $M$
number of pairs, for which ($\tau$ - $\Delta \tau$/2) $\leq$ $\Delta t_{ij}$ $<$ ($\tau$ + $\Delta \tau$/2), we obtain,
\begin{equation}
 DCF(\tau) = \frac{1}{M} UDCF_{ij},
\end{equation}
and the error of DCF is defined as,
\begin{equation}
 \sigma_{DCF}(\tau) = \frac{1}{M-1} \Bigg\{\sum[UDCF_{ij} - DCF(\tau)] \Bigg\}^{1/2}\,.
\end{equation}
All the above steps are incorporated within the python script \texttt{PyDCF} (\citealt{Robertson2015}) which we have used for the time lag estimation.

\subsubsection{$z$-transformed DCF}

The $z$-transformed discrete correlation function (DCF) improves the classical DCF by replacing the equal time bins by equal population bins and applies the Fisher's $z$ transformation to stabilize the skewed distribution of the cross-correlation function \citep{1997ASSL..218..163A}. In this way, the zDCF outperforms DCF especially for undersampled, sparse, and heterogeneous datasets. The minimum number of light curve points per bin, for which the correlation coefficient is estimated, can be set and we specify it in the main text. In addition, the uncertainty of the cross-correlation function as well as of the time-delay is estimated via a specified number of Monte Carlo simulations. The uncertainty of the candidate time-delay peak is calculated using the Maximum-Likelihood function based on the zDCF values.

\onecolumn
\begin{center}
\begin{longtable}{c|c|c|c}
\caption{Instruments are: 1 - OGLE, 2 - SALT, 3 - BMT, 4 - SSO, 5 - lesedi, 6 - CTIO, 7 - SAAO.}
\label{tab:photometry} \\ 
\hline
\hline
JD & magnitude (V-band) & Error & Instrument\\
-2 450 000 & [mag] & [mag] & No. \\
\hline
6199.799 & 16.954 & 0.005 & 1 \\
6210.817 & 16.960 & 0.004 & 1 \\
6226.679 & 16.943 & 0.005 & 1 \\
6246.698 & 16.945 & 0.004 & 1 \\
6257.750 & 16.958 & 0.006 & 1 \\
6268.683 & 16.962 & 0.004 & 1 \\
6277.685 & 16.972 & 0.003 & 1 \\
6286.669 & 16.984 & 0.005 & 1 \\
6297.618 & 17.005 & 0.004 & 1 \\
6307.576 & 17.014 & 0.004 & 1 \\
6317.643 & 16.990 & 0.005 & 1 \\
6330.658 & 17.022 & 0.004 & 1 \\
6351.550 & 17.046 & 0.005 & 1 \\
6363.575 & 17.050 & 0.004 & 1 \\
6379.488 & 17.051 & 0.005 & 1 \\
6379.496 & 17.045 & 0.005 & 1 \\
6387.514 & 17.065 & 0.004 & 1 \\
6637.672 & 17.154 & 0.004 & 1 \\
6651.623 & 17.163 & 0.004 & 1 \\
6665.606 & 17.167 & 0.004 & 1 \\
6678.601 & 17.159 & 0.004 & 1 \\
6689.675 & 17.136 & 0.004 & 1 \\
6700.638 & 17.145 & 0.006 & 1 \\
6715.578 & 17.117 & 0.004 & 1 \\
6740.493 & 17.102 & 0.004 & 1 \\
7015.536 & 17.013 & 0.012 & 2 \\
7036.654 & 17.024 & 0.004 & 1 \\
7048.656 & 17.021 & 0.004 & 1 \\
7060.607 & 17.031 & 0.005 & 1 \\
7084.538 & 17.052 & 0.005 & 1 \\
7110.248 & 17.066 & 0.013 & 2 \\
7118.510 & 17.055 & 0.005 & 1 \\
7240.633 & 17.056 & 0.012 & 2 \\
7253.895 & 17.058 & 0.004 & 1 \\
7261.886 & 17.020 & 0.004 & 1 \\
7267.918 & 17.021 & 0.005 & 1 \\
7273.850 & 17.058 & 0.004 & 1 \\
7295.846 & 17.052 & 0.005 & 1 \\
7306.784 & 17.082 & 0.004 & 1 \\
7317.743 & 17.101 & 0.005 & 1 \\
7327.778 & 17.109 & 0.005 & 1 \\
7340.709 & 17.126 & 0.004 & 1 \\
7343.359 & 17.132 & 0.012 & 2 \\
7355.698 & 17.119 & 0.005 & 1 \\
7363.669 & 17.109 & 0.004 & 1 \\
7374.712 & 17.138 & 0.004 & 1 \\
7385.561 & 17.154 & 0.004 & 1 \\
7398.621 & 17.145 & 0.004 & 1 \\
7415.589 & 17.149 & 0.004 & 1 \\
7423.396 & 17.112 & 0.012 & 2 \\
7426.570 & 17.135 & 0.004 & 1 \\
7436.529 & 17.123 & 0.005 & 1 \\
7447.531 & 17.115 & 0.004 & 1 \\
7457.526 & 17.140 & 0.004 & 1 \\
7665.464 & 17.126 & 0.012 & 2 \\
7688.436 & 17.107 & 0.012 & 2 \\
7717.708 & 17.106 & 0.004 & 1 \\
7807.340 & 17.073 & 0.012 & 2 \\
7968.647 & 17.124 & 0.011 & 2 \\
8041.431 & 17.170 & 0.012 & 2 \\
8090.742 & 17.195 & 0.036 & 3 \\
8091.789 & 17.140 & 0.041 & 3 \\
8092.738 & 17.139 & 0.040 & 3 \\
8096.809 & 17.138 & 0.028 & 3 \\
8098.812 & 17.145 & 0.028 & 3 \\
8100.531 & 17.151 & 0.012 & 2 \\
8127.660 & 17.109 & 0.029 & 3 \\
8134.621 & 17.100 & 0.028 & 3 \\
8141.637 & 17.121 & 0.028 & 3 \\
8165.516 & 17.101 & 0.029 & 3 \\
8173.512 & 17.111 & 0.029 & 3 \\
8180.504 & 17.117 & 0.029 & 3 \\
8196.539 & 17.087 & 0.028 & 3 \\
8205.520 & 17.097 & 0.030 & 3 \\
8210.504 & 17.050 & 0.030 & 3 \\
8367.887 & 17.042 & 0.032 & 3 \\
8375.529 & 17.033 & 0.012 & 2 \\
8414.754 & 17.029 & 0.033 & 3 \\
8434.360 & 17.017 & 0.012 & 2 \\
8463.544 & 16.995 & 0.011 & 2 \\
8498.446 & 16.983 & 0.012 & 2 \\
8531.121 & 16.952 & 0.035 & 3 \\
8532.105 & 16.958 & 0.035 & 3 \\
8539.102 & 16.969 & 0.034 & 3 \\
8542.039 & 16.981 & 0.033 & 3 \\
8558.059 & 16.964 & 0.036 & 3 \\
8569.992 & 17.007 & 0.033 & 3 \\
8578.984 & 16.971 & 0.034 & 3 \\
8724.569 & 16.973 & 0.012 & 2 \\
8821.302 & 17.059 & 0.012 & 2 \\
8852.467 & 17.070 & 0.012 & 2 \\
8884.102 & 17.062 & 0.034 & 3 \\
8901.059 & 17.043 & 0.034 & 3 \\
9075.609 & 17.103 & 0.012 & 2 \\
9090.672 & 17.043 & 0.015 & 4 \\
9099.988 & 17.007 & 0.016 & 5 \\
9110.766 & 17.042 & 0.014 & 4 \\
9116.499 & 17.029 & 0.011 & 2 \\
9120.957 & 17.036 & 0.017 & 5 \\
9136.012 & 17.015 & 0.016 & 5 \\
9155.902 & 17.038 & 0.020 & 7 \\
9172.824 & 17.044 & 0.015 & 7 \\
9181.047 & 17.034 & 0.017 & 6 \\
9194.051 & 16.902 & 0.049 & 5 \\
9196.969 & 17.022 & 0.016 & 5 \\
9207.879 & 17.053 & 0.015 & 7 \\
9209.090 & 17.070 & 0.015 & 6 \\
9224.211 & 17.072 & 0.009 & 6 \\
9235.435 & 17.090 & 0.012 & 2 \\
9243.809 & 17.060 & 0.015 & 5 \\
9258.957 & 17.066 & 0.015 & 5 \\
9291.273 & 17.110 & 0.012 & 2 \\
9298.266 & 17.143 & 0.013 & 2 \\
\hline

\end{longtable}
\end{center}
\twocolumn

\end{appendix}
\end{document}